\documentclass[12pt]{article}
\usepackage{amsfonts, amsmath, graphics}
\makeatletter
\@addtoreset{figure}{section}
\def\thefigure{\thesection.\@arabic\c@figure}
\def\fps@figure{h, t}
\@addtoreset{table}{bsection}
\def\thetable{\thesection.\@arabic\c@table}
\def\fps@table{h, t}
\@addtoreset{equation}{section}

\makeatother

\begin{document}


\title{\vskip -0.6in
Integrable vs Nonintegrable \\
Geodesic Soliton Behavior
}

\author{Oliver B. Fringer$^{1,2}$ and Darryl D. Holm$^{2}$}

\date{\today}

\maketitle

\footnotesize
${}^{1}$Dept. of Civil
    and Environmental Engineering, Stanford University, Stanford, CA
    94305.  email: fringer@leland.stanford.edu,
    www: http://rossby.stanford.edu/$^\sim$fringer

  ${}^{2}$Theoretical Division and Center for Nonlinear Studies, Los
    Alamos National Laboratory, Los Alamos, NM 87545.
    email: dholm@lanl.gov,  www: http://cnls.lanl.gov/$^\sim$dholm
\normalsize

\begin{abstract}
We study confined solutions of certain evolutionary partial differential
equations (pde) in $1+1$ space-time. The pde we study are Lie-Poisson
Hamiltonian systems for quadratic Hamiltonians defined on the dual of the Lie
algebra of vector fields on the real line. These systems are also
Euler-Poincar\'e equations for geodesic motion on the diffeomorphism group in
the sense of the Arnold program for ideal fluids, but where the kinetic
energy metric is different from the $L^2$ norm of the velocity. These pde
possess a finite-dimensional invariant manifold of particle-like
(measure-valued) solutions we call ``pulsons.'' We solve the particle dynamics
of the two-pulson interaction analytically as a canonical Hamiltonian system
for geodesic motion with two degrees of freedom and a conserved momentum. The
result of this two-pulson interaction for rear-end collisions is elastic
scattering with a phase shift, as occurs with solitons. In contrast, head-on 
antisymmetric collisons of pulsons tend to form singularities.
\medskip

Numerical simulations of these pde show that their evolution by geodesic
dynamics for confined (or compact) initial conditions in various
nonintegrable cases possesses the same type of multi-soliton behavior
(elastic collisons, asymptotic sorting by pulse height) as the
corresponding integrable cases do. We conjecture this behavior occurs
because the integrable two-pulson interactions dominate the dynamics on the
invariant pulson manifold, and this dynamics dominates the pde initial value
problem for most choices of confined pulses and initial conditions of finite
extent.
\end{abstract}
\clearpage
\tableofcontents


\section{Introduction}
We study particle-like dynamics on a set of finite-dimensional invariant
manifolds for certain measure-valued solutions, called ``pulsons,'' of the
family of evolutionary integral partial differential equations given by
\begin{equation} \label{EPeqn}
m_t + um_x + 2 mu_x = 0\,,
\quad
\lim_{|x|\to\infty}m=0\,.
\end{equation}
Here subscripts denote partial derivatives,
$m:\mathbb{R}\times\mathbb{R}\rightarrow\mathbb{R}$ is a
real, measure-valued map on $1+1$ space-time with coordinates $x,t$, and
the function $u$ is defined by the convolution integral,
\begin{equation} \label{Ueqn}
u(x,t) = \int_{-\infty}^{\infty} g(x-y)\,m(y,t)\,dy =: g \ast m\,.
\end{equation}
The integral kernel, or Green's function, $g(x)$, is taken to be even,
$g(-x)=g(x)$, of confined spatial extent, and such that the quadratic
integral quantity
\begin{equation} \label{Hdef}
H = \frac{1}{2} \int_{-\infty}^{\infty}
m\,g \ast m\,dx
\quad\hbox{with}\quad
\frac{\delta H}{\delta m}=u\,,
\end{equation}
is positive definite; so that $H$ defines a norm. Physically, $m$
is a momentum density, associated to a velocity distribution
$u=g \ast m$, and $H$ is the kinetic energy for the dynamics. Evenness of
the Green's function $g(x)$ implies that equation (\ref{EPeqn}) conserves
the kinetic energy $H$ and the total momentum
$P=\int_{-\infty}^{\infty} m\, dx$, for solutions $m$ of equation
(\ref{EPeqn}) that vanish at spatial infinity. The ``mass'' $M =
\int_{-\infty}^{\infty} \sqrt{m}\, dx$ is also conserved for such solutions.

In a certain formal sense, equation (\ref{EPeqn}) is hyperbolic, as it
follows from a pair of equations of hydrodynamic type, namely
\begin{equation} \label{Hyp_eqn}
\ell_t + u\ell_x = 0\,,\quad \eta_t + (\eta u)_x = 0\,,\quad
\hbox{with} \quad u=-g \ast (\eta \ell_x)\,.
\end{equation}
This pair of equations implies equation (\ref{EPeqn}), upon
setting $m=-\eta\ell_x$. Note that $\ell$ is preserved along flow lines of $u$
and the integrated ``mass'' $\int\eta dx$ is conserved for solutions of
equations
(\ref{Hyp_eqn}) for which $u$ vanishes at spatial infinity. Note also that
$\ell_x$ satisfies the same equation as $\eta$ does.  Thus, these equations
preserve the relation $\eta=\pm\ell_x$, provided it holds initially.  Hence,
equation (\ref{EPeqn}) preserves sign-definitiveness of $m$, provided $m$
can be
initially expressed as $m = \pm \eta^2$, for some function
$\eta(x,0)$ on the real line. Moreover, $\eta^2$ also satisfies equation
(\ref{EPeqn}), and this equation is the second in a hierarchy of equations
implied by (\ref{Hyp_eqn}) for the powers of $\eta$. Because of its special
geometric properties, we shall concentrate our attention here on the
family of equations (\ref{EPeqn}) for various choices of the Green's function,
$g(x)$.

\paragraph{Hamiltonian and geodesic properties.}
Equation (\ref{EPeqn}) is expressible in Lie-Poisson Hamiltonian
form $m_t=\{m,H\}$ with Hamiltonian $H$ given in equation (\ref{Hdef}) and
Lie-Poisson bracket defined by
\begin{equation} \label{LPBracketeqn}
\{F,H\} = -\int_{-\infty}^{\infty}
m\bigg[\frac{\delta F}{\delta m},\frac{\delta H}{\delta m}\bigg]\, dx
= -\int_{-\infty}^{\infty} \frac{\delta F}{\delta m}\,
(\partial m + m \partial)\, \frac{\delta H}{\delta m}\,dx \,,
\end{equation}
where we have integrated by parts and $\partial$ denotes the operator
$\partial/\partial{x}$. This Lie-Poisson bracket is defined on
$\mathfrak{g}^{\ast}$, the dual of the Lie algebra
$\mathfrak{g}$ of vector fields on the real line that vanish at spatial
infinity and possess the Lie bracket operation
$ad:\mathfrak{g}\times\mathfrak{g}\to\mathfrak{g}$ defined by
$ad_fh=[f,h]=fh_x-hf_x$, for $f,h\in\mathfrak{g}$. Thus, equation
(\ref{EPeqn}) may be rewritten for $m\in\mathfrak{g}^{\ast}$ as
\begin{equation} \label{LPeqn}
m_t = \{m,H\}
= -(\partial{m}+m\partial)\frac{\delta H}{\delta m}
= -ad^{\ast}_{{\delta H}/{\delta m}}m\,,
\end{equation}
where $ad^{\ast}$ is the operation on $\mathfrak{g}^{\ast}$ dual to
the $ad$-operation on $\mathfrak{g}$.
By Legendre transforming from $\mathfrak{g}^{\ast}$ to $\mathfrak{g}$
using the relation ${\delta H}/{\delta m}=u\in\mathfrak{g}$, equation
(\ref{LPeqn}) may be re-expressed as an Euler-Poincar\'e
equation~\cite{HP[1901]},~\cite{Arnold[1966]},~\cite{MR[1994]}
\begin{equation} \label{EPeqn1}
\frac{\partial}{\partial t} \frac{\delta L}{\delta u}
= -ad^{\ast}_u\, \frac{\delta L}{\delta u}\,,
\end{equation}
for the Lagrangian
\begin{equation} \label{Lag}
L = \int_{-\infty}^{\infty} mu\,dx - H
  = \frac{1}{2} \int_{-\infty}^{\infty} u\,Q\,u\,dx \,,
\end{equation}
where $g(x)$ is the Green's function for the self-adjoint
positive operator $Q$, i.e.,
\begin{equation} \label{green-def}
Q\,g(x)=2\delta(x)\,,
\end{equation}
with Dirac measure $\delta(x)$. Thus, $m={\delta L}/{\delta u}=Qu$ is the
momentum density conjugate to the velocity $u\in\mathfrak{g}$, and $u$
satisfies the Euler-Poincar\'e equation,
\begin{equation} \label{EPeqn2}
\frac{\partial}{\partial t} Qu
= -ad^{\ast}_u\, Qu\,.
\end{equation}
Replacing $m$ by $Qu$ in (\ref{EPeqn}) reveals that this equation is
not Galilean invariant (i.e., it changes form under $x\to x+ct$, $u\to
u+c$, $t\to t$) unless $m\to m$ under such a transformation.
Typically, $m\to m+\kappa$, with $\kappa$ a constant depending on
$c$ under a Galilean transformation, so that equation
(\ref{EPeqn}) becomes
\begin{equation} \label{EP_kappa_eqn}
m_t + um_x + 2mu_x + 2\kappa u_x = 0\,,
\end{equation}
which for $\kappa \ne 0$ introduces linear dispersion.  Thus, equation
(\ref{EPeqn})
should best be considered as a family of equations parameterized by
the function $g(x)$ and the constant $\kappa$, here taken as $\kappa =
0$.  The aim of this paper is to explore the solution behavior of this
family of equations for the initial value problem, under variations of
$g(x)$ and the initial conditions.  The effects of $\kappa \ne 0$ will
also be discussed briefly in the Appendix.

Equation (\ref{EPeqn2}) is formally the equation for geodesic motion on the
diffeomorphism group with respect to the metric given by the Lagrangian $L$ in
equation (\ref{Lag}), which is right-invariant under the action of the
diffeomorphism group. See Holm, Marsden and Ratiu~\cite{HMR[1998a]} for
detailed discussions, applications and references to Euler-Poincar\'e equations
of this type for ideal fluids and plasmas. See Ovsienko and
Khesin~\cite{O&K[1987]} and Segal~\cite{Segal[1991]} for discussions of the
Korteweg-de Vries (KdV) equation from the viewpoint of the Euler-Poincar\'e
theory of geodesic motion on the Bott-Virosoro group.
Misiolek~\cite{Misiolek[1997]} has given a similar interpretation  to equation
(\ref{EPeqn}) in the integrable case,
$g(x)=e^{-|x|}$~\cite{CH[1993]},~\cite{CHH[1994]}. See also Holm et
al.~\cite{HMR[1998a]},~\cite{HMR[1998b]} for generalizations of
equation (\ref{EPeqn}) to higher spatial dimensions, and
Shkoller~\cite{S1[1998]} and Holm et al.~\cite{HKMRS[1998]} for
generalizations to
Riemannian manifolds.  Some of the most interesting solutions of the
systems we study actually leave the diffeomorphism group due to a loss of
regularity.  Such solutions must be interpreted in the sense of
generalized flows, as done by Brenier~\cite{B[1998]} and
Shnirelman~\cite{S2[1998]}.
The functional-analytic study of these solutions is made in Holm et
al.~\cite{HMS[1998]}.  In this paper we shall formally consider such
solutions to represent geodesic motion and hence keep the above nomenclature.
These dynamics can be formulated either in a periodic
domain, or on the real line. For the analysis here we shall work on the
real line. The numerics will be conducted in a periodic domain.

\paragraph{Completely integrable cases.} The Hamiltonian system
(\ref{EPeqn}) is known to be completely integrable in the
Liouville-Arnold sense for three cases:
$g = \delta(x)$; $g = 1-|x|$ on $|x|<1$ (and $g=0$ elsewhere); and $g =
e^{-|x|}$. These cases are identified with the following Lie-Poisson
equations and solution behavior.
\begin{equation} \label{integ.g}
\begin{array}{lll}
g =  \delta(x)\,, & m_t + 3mm_x = 0
&\hbox{shocks},\\ \\
g = \left\{
  \begin{array}{ll}
  1-|x| & \hbox{for}\, |x|< 1,\\
  0 & \hbox{for}\, |x|\ge1,
  \end{array} \right.
&(u_t+uu_x)_{xx}=\frac{1}{2}(u_x^2)_x
&\hspace{-0.3in}
\hbox{compactons},\\ \\
g = e^{-|x|}\,, & \hspace{-0.5in}
(1-\partial^2)(u_t+uu_x) = -\,(u^2 + \frac{1}{2}u_x^2)_x
&\hbox{peakons}.
\end{array}
\end{equation}
The first case yields the Riemann equation, which governs the formation of
shocks. The second case yields a Galilean invariant equation in the Harry Dym
hierarchy (at the KdV shallow water position) which describes the propagation
of weakly nonlinear orientation waves in a massive nematic liquid crystal
director field~\cite{H&Z[1994]}. The compacton solutions of this equation are
triangular pulses with compact support that interact as solitons.  The third
case, $g = e^{-|x|}$, yields the Camassa-Holm equation, whose peakon solutions
describe a limiting situation of unidirectional shallow water dynamics
\cite{CH[1993]},~\cite{CHH[1994]}. Integrable compactons and peakons are
discussed further in~\cite{CH[1993]},~\cite{CHH[1994]},~\cite{H&Z[1994]} for
motion on the real line, and in~\cite{ACHM[1994]},~\cite{ACHM[1995]} and
references therein for the periodic case. See also
references~\cite{C&M[1998]},~\cite{V&L[1997]} for other recent discussions
of the Camassa-Holm equation. Note that, from (\ref{integ.g}), the Dym
shallow water equation for compactons arises as a ``high wavenumber limit''
of the Camassa-Holm equation for peakons. The latter two integrable cases
each have an associated isospectral problem whose discrete spectrum
determines the asymptotic speeds of the solitons that emerge from their
initial conditions. Here we shall use these integrable solutions as
comparisons in assessing the nonintegrable, but \textbf{qualitatively
identical} pulse interaction behavior for other choices of $g(x)$,
which turns out to be the pulse profile.

\paragraph{Traveling waves.} Traveling wave solutions of equation
(\ref{EPeqn}) of the form $m(x-ct)$, with $u(x-ct)= g \ast m$, satisfy
\begin{equation}
-(c-u)m' + 2mu' = 0\,,
\end{equation}
whose first integral is
\begin{equation}
(c-u)^2m = a^2 = \hbox{const}.
\end{equation}
These traveling waves are critical points of the sum of conserved
quantities, $H-cP+2aM$.  The boundary condition $\lim_{|x|\to \infty}m
= 0$ implies that the constant $a$ must vanish.  Therefore, $m$ must
vanish except where $g\ast m = u = c$.  Thus, equation (\ref{EPeqn}) admits
Dirac measure valued (particle-like) traveling wave solutions $m =
2c\delta(x-ct)$, for which $u=cg(x-ct)$. Substitution of these formulae for
$m$ and $u$  into (\ref{EPeqn})
and integration by parts against a smooth test function (using $g'(0) =
0$ from evenness of $g$) verifies this solution, provided $g(0)=1$.
Therefore, the traveling wave's speed $c$ is equal to the peak height of
its velocity profile, and the Green's function $g(x-ct)$ is the
normalized shape of this profile. We shall
study the interaction dynamics of a superposition of ``pulsons'' in the
following form
\begin{equation} \label{M-puls.def}
m(x,t) = \sum_{i=1}^{N} 2p_i (t)\delta (x-q_i (t)\,)\,,
\end{equation}
for which the velocity superposes as~\cite{footnote}
\begin{equation} \label{U-puls.def}
u(x,t) = \sum_{i=1}^{N} p_i (t)g(x-q_i (t)\,)\,,
\end{equation}
and the Hamiltonian $H$ in (\ref{Hdef}) becomes (up to an unimportant
factor of 2)
\begin{equation} \label{H_N-def}
H_N = \frac{1}{2}\sum_{i,j=1}^{N} p_i p_j\, g(q_i-q_j)\,.
\end{equation}
Substitution of these superpositions of particle-like solutions into
(\ref{EPeqn}) shows that they form an invariant manifold under the
Lie-Poisson dynamics of (\ref{EPeqn}), provided the parameters $(p_i,q_i)$,
$i=1,\dots,N$, satisfy the finite-dimensional particle dynamics equations,
\begin{equation} \label{ODEeqn}
\dot{p}_i = -p_i\sum_{j=1}^{N} p_j\, g'(q_i-q_j)\,,
\qquad
\dot{q}_i = \sum_{j=1}^{N} p_j\, g(q_i-q_j)
\,.
\end{equation}
These are precisely Hamilton's canonical equations for the ``collective''
Hamiltonian $H_N$ in (\ref{H_N-def}). These equations describe geodesic
motion on an $N$-dimensional surface with coordinates $q_i$,
$i=1,\dots,N$, and co-metric $g^{ij}=g(q_i-q_j)$. Equations (\ref{ODEeqn})
are integrable for any finite $N$ for two
cases~\cite{Calogero[1995]},~\cite{C&F[1996]}; namely,
$g(x)=\lambda+\mu\cos(\nu{x})+\mu'\sin(\nu|x|)$ and
$g(x)=\lambda+\alpha|x|+\beta{x^2}$, where
$\{\lambda,\mu,\mu',\nu,\alpha,\beta\}$ are free constants. These cases
correspond to the pde for compactons and peakons in (\ref{integ.g}) known to
be integrable by the isospectral method, when $g(x)$ is required to be
spatially confined. We shall choose Green's functions $g$ that are spatially
confined (or have compact support); so that $g(x)$ and $g{\,}'(x)$ are
negligible for $|x|>D$, with $D$ the interaction range between particles. Thus,
once the separations between every pair of peaks satisfies $|q_i-q_j|>D$,
equations (\ref{ODEeqn}) will become essentially $\dot{p}_i=0$ and
$\dot{q}_i=p_i$. Subsequently, the peak positions of the separated pulsons will
undergo free linear motion $p_i=c_i$, $q_i=c_it+q_i^0$ for a set of $2N$
constants, the speeds $c_i$ and phases $q_i^0$, for $i=1,\dots,N$. Thus, the
peaks will separate linearly in time, in proportion to their difference in
heights, and become linearly ordered according to height, with the faster ones
to the right.

Numerically, the finite-dimensional invariant manifold of superposed
solutions given in (\ref{M-puls.def}), (\ref{U-puls.def}) and satisfying
(\ref{ODEeqn}) shall be shown for various confined initial velocity
distributions and choices of pulse shape $g(x)$ to describe stable pulses
which interact elastically like solitons do, sort themselves according to
height (as expected for geodesic motion), and dominate the solution of the
initial value problem via their two-body interactions.

\section{Pulson-Pulson and Pulson-Antipulson Interactions}
Before we show numerical solutions of initial value problems for
various choices of $g(x)$, we shall present the solution of the 2-pulson and
pulson-antipulson interaction equations for arbitrary $g(x)$. We
follow~\cite{CH[1993]},~\cite{CHH[1994]} for the analytical solution of the
two-peakon and peakon-antipeakon interaction dynamics for the Camassa-Holm
equation. For $N=2$, the collective Hamiltonian (\ref{H_N-def}) becomes
\begin{equation} \label{2PHdef}
H = \frac{1}{2}(p^2_1 + p^2_2) + p_1 p_2 g(q_2 - q_1)\,.
\end{equation}
Defining sum and difference canonical variables as
$P = p_1 + p_2$, $Q = q_1 + q_2$, and $p = p_2 - p_1$,
$q = q_2 - q_1 $, respectively, transforms this Hamiltonian to
\begin{equation} \label{2PHdif}
H = \frac{1}{2}P^2 - \frac{1}{4}(P^2-p^2)(1-g(q))\,,
\end{equation}
which is independent of the sum coordinate $Q$, so that $P$ is
conserved. Thus $H$ and $P$ are both
constants of motion. Hence, equation (\ref{2PHdif}) relates the phase
space coordinates $(p,q)$ by
\begin{equation} \label{p_vs_q_eqn*}
p^2 = \frac{4H-2P^2}{1 - g(q)}+P^2
=\frac{-4c_1c_2}{1 - g(q)}+(c_1+c_2)^2\,,
\end{equation}
upon writing the values of the constants $H$ and $P$ as
$H=\frac{1}{2}(c_1^2+c_2^2)$ and $P=(c_1+c_2)$, for asymptotic speeds $c_1$
and $c_2$. In this relation, the condition $g(0)=1$ for $q=0$ will produce
singular
behavior for $p$, should the pulsons overlap. For
2-pulson, or 2-antipulson collisons, we have $c_1c_2>0$, so the peak
separation $q$ cannot vanish in these cases for real $p$. However, for
pulson-antipulson collisons, we have $c_1c_2<0$; so $q=0$ may occur and
the relative momentum $p$ diverges when this happens.

\begin{figure}[ht!]
\centerline{
\scalebox{.5}{\includegraphics{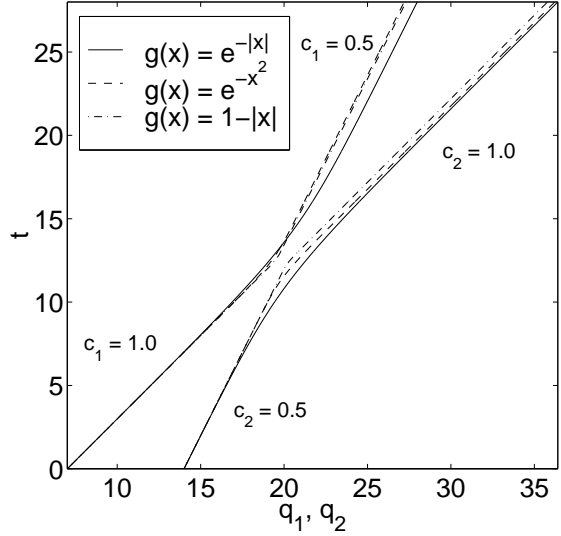}}
}
\caption{Space-time trajectories $q_1(t)$ and $q_2(t)$ for three typical
Green's
  functions.  For any spatially confined Green's function, these
  space-time plots become linear, asymptotically in time.  Note that the
relative
  separation $q = q_2-q_1$ remains positive, so the particles retain their
  order.}
\label{q-vs-t_fig}
\end{figure}

\begin{figure}[ht!]
\centerline{
\scalebox{.5}{\includegraphics{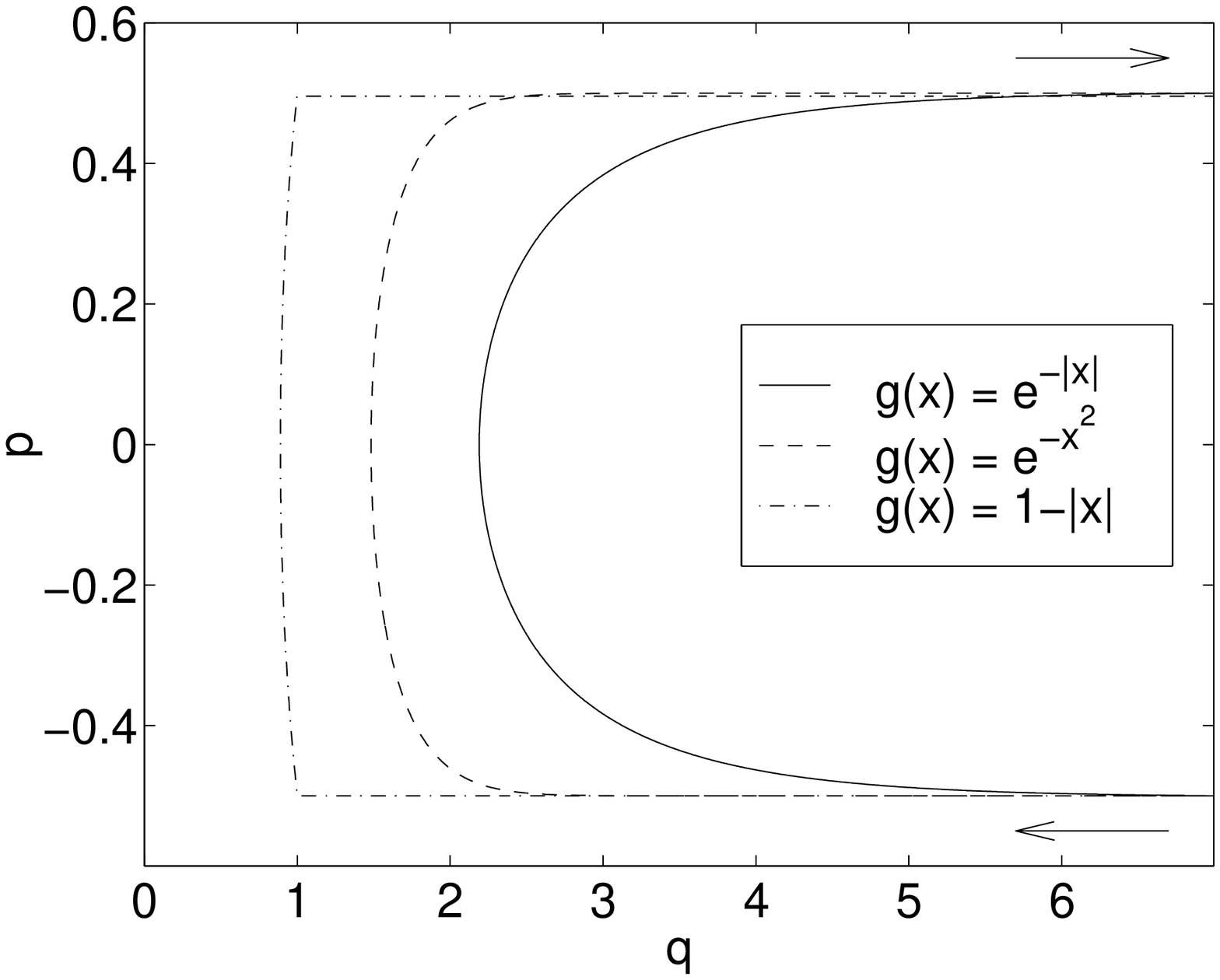}}
}
\caption{Phase space trajectories $p(t)$ vs $q(t)$ of rear-end collisons
for three typical
  Green's functions.  For rear-end collisons these trajectories are
nonsingular. Note that
  $p = p_2-p_1$ changes sign, while $q = q_2-q_1$ remains positive;
  so the particles exchange their momentum, but retain their order.}
\label{q-vs-p_fig}
\end{figure}

The sum and difference variables obey the canonical equations
\begin{equation} \label{can-odes}
{\begin{array}{ll}
\dot{P} = -2 \frac{\partial H}{\partial Q} = 0\,, &
\dot{Q} =  2 \frac{\partial H}{\partial P} = P(1+g(q))\,, \\
\dot{p} = -2 \frac{\partial H}{\partial q} =
-\frac{1}{2}(P^2-p^2)g'(q)\,, &
\dot{q} =  2 \frac{\partial H}{\partial p} = p(1-g(q))\,.
\end{array}}
\end{equation}
Conservation of the total momentum $P=\sum_{i=1}^N p_i$ holds for
arbitrary $N$, and for the case $N=2$ it is sufficient for solvability of
the dynamics.
Substituting $p$ from (\ref{p_vs_q_eqn*}) into the $\dot{q}$
equation in (\ref{can-odes}) gives a quadrature formula for the dynamics of
the separation between peaks, $q(t)$.  Namely,
\begin{equation} \label{quadrat-1}
\pm (t - t_0) = \int_{q(t_0)}^{q(t)} \frac{dq'}
{\left[P^2 g^2(q') - 4 H g(q') + \left( 4 H - P^2 \right)
 \right]^{1/2}}\,.
\end{equation}
Seen as a collision between two initially well-separated ``particles'' with
initial speeds $c_1$ and $c_2$, the separation $q(t)$ reaches a
nonzero distance of closest approach for 2-pulson collisions with $c_1c_2>0$
and this quadrature formula produces only a phase shift (a delay or
advance of position relative to the extension of the incoming
trajectory, without a change in asymptotic speed) of the two
asymptotically linear $q(t)$ trajectories. These trajectories are shown for
several choices of $g(x)$ in Figure \ref{q-vs-t_fig}.
The phase space trajectories of rear-end collisions for three typical Green's
functions are shown in Figure \ref{q-vs-p_fig}. Notice that $p$ is
nonsingular and $q$ remains positive throughout, so the particles retain their
order. The momentum transfer in these rear-end collisions occurs rather
suddenly
over a small range of separation distance $q$, especially for wave forms
with compact
support.

\begin{figure}[ht!]
\centerline{
\scalebox{.5}{\includegraphics{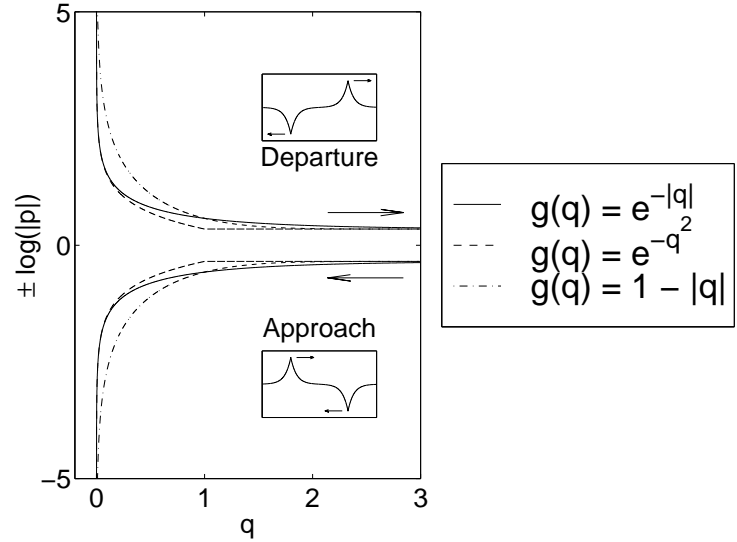}}
}
\caption{Phase space trajectories of head-on collisions for three typical
Green's
  functions.  As $q\to0^+$ at the collison point, $1/p$ passes through
zero. Note that
  $q$ remains positive throughout the collision, so the particles retain their
  order. In particular, they do not ``pass through'' each other.}
\label{pulson_phase_fig}
\end{figure}

\paragraph{Head-on pulson-antipulson collision.}
For the special case of completely antisymmetric pulson-antipulson
collisions, for which
$p_1 = -p_2 = -p/2$ and $q_1 = -q_2 = -q/2$ (so that $P=0$ and $Q=0$) the
quadrature
formula (\ref{quadrat-1}) reduces to
\begin{equation} \label{quadrat-2}
\pm (t-t_0) = \frac{1}{\sqrt{4 H}}
\int_{q(t_0)}^{q(t)} \frac{dq'}{\sqrt{1-g(q')}}\,.
\end{equation}
For this case, the conserved Hamiltonian (\ref{2PHdif}) is
\begin{equation} \label{simp_H_eqn}
H = \frac{p^2}{4}[1-g(q)]\,.
\end{equation}
After the collision, the pulson and antipulson separate and travel
oppositely apart; so that asymptotically in time $g(q)\to0$, $p\to2c$, and
$H\to{c}^2$,
where $c$ (or $-c$) is the asymptotic speed (and amplitude) of the pulson (or
antipulson).  Setting $H={c}^2$ in equation (\ref{simp_H_eqn}) gives a
relation for the pulson-antipulson $(p,q)$ phase trajectories for any Green's
function,
\begin{equation} \label{p_vs_q_eqn}
p = \mp \frac{2 c}{\sqrt{1 - g(q)}}\,.
\end{equation}
Notice that $1/p$ passes through zero when $q\to0$, since $g(0) = 1$. The
relative
momentum $p$, initially at $-2c$, diverges to $-\infty$ at the ``bounce''
collision point
as $q\to 0^+$. Then $p$ changes sign, $q$ increases again and $p$
asymptotes to $2c$ from
above. Thus, $p$ diverges as $q\to0^+$ and switches branches of the square
root, from
negative to positive.  Note also that $q>0$ throughout, so the particles
retain their
order. The phase space trajectories of head-on collisions for three typical
Green's
functions are shown in Figure \ref{pulson_phase_fig}.

In terms of relative momentum and separation, $p$ and $q$, the solution
(\ref{U-puls.def})
for the velocity $u$ in the head-on pulson-antipulson collision becomes
\begin{equation}
u(x,t) = \frac{-p}{2}g(x+q/2) + \frac{p}{2}g(x-q/2)\,.
\end{equation}
Using equation (\ref{p_vs_q_eqn}) to eliminate $p$ allows this solution
to be written as a function of position $x$ and the separation
between the pulses $q$ for {\it any} pulse shape $g(x)$ as
\begin{equation} \label{vel_collide_eqn}
u(x,q) = \frac{\mp{c}}{\sqrt{1-g(q)}}\Big[g(x+q/2) - g(x-q/2)\Big],
\end{equation}
where $c$ is the pulson speed at sufficiently large separation and the
dynamics of the separation $q(t)$ is given implicitly by equation
(\ref{quadrat-2}) with $\sqrt{4H}=2c$. Equation (\ref{vel_collide_eqn}) is the
exact analytical solution for the pulson-antipulson collision.

\section{Numerical Simulations}

We shall summarize our choices of Green's functions and briefly discuss
the numerical method used to compare their pde interaction dynamics in
numerical simulations. These numerical simulations demonstrate the
qualitative similarity in solution behavior for these Green's functions in
both rear-end (2-pulson) and head-on (pulson-antipulson) collisions. The
numerical results show that the dynamics of the initial value problem is
dominated by Green's function wave forms that collide elastically with each
other and sort themselves according to height, as solitons do. The
reversibility of the
numerical dynamics for the initial value problem is also shown, by
demonstrating that the dynamics of a set of separated peaked solitary waves
may be reversed to reconstruct a smooth initial distribution.

\subsection{Choice of Green's Functions}

\begin{figure}[ht!]
\centerline{
\scalebox{.4}{\includegraphics{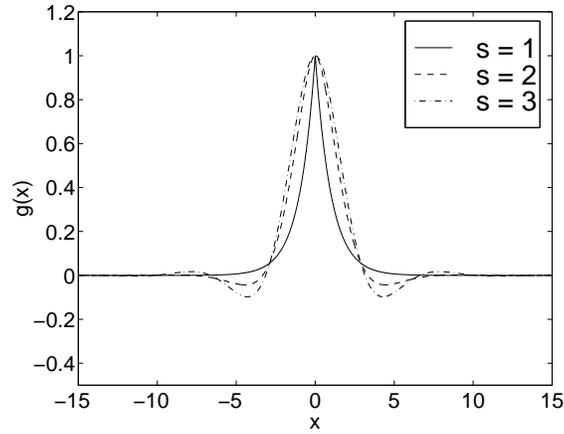}}
}
\caption{Pulson shapes for s=1,2, and 3.}
\label{s123_pulsons_fig}
\end{figure}

The choice $g(x) = e^{-|x|}$ satisfies
$(1 - \partial^2)g(x) = 2 \delta (x)$, for which the Hamiltonian (\ref{Hdef})
becomes
\begin{equation} \label{H1eqn}
H_1 = \frac{1}{2}\int mu\,dx = \frac{1}{2}\int (u^2 + u^2_x) dx\,.
\end{equation}
This is the $H_1$ norm which generates the peakon dynamics studied
in~\cite{CH[1993]},~\cite{CHH[1994]},~\cite{ACHM[1994]},~\cite{ACHM[1995]}.
It is natural to consider higher-derivative norms of this type, such as
\begin{equation} \label{Hseqn}
H_s = \frac{1}{2}\int (u^2 + u^2_s) dx\,,
\end{equation}
with $u_s = \partial^s{u}/\partial{x^s}$. The corresponding normalized
Green's function
\begin{equation} \label{Greenseqn}
g^{(s)}(x) = \frac{1}{R}\sum_{j=0}^{s-1} r_j e^{-r_j |x|}\,,
\end{equation}
with $r_j = exp \left ( \frac{i \pi}{s} \left (j - \frac{s-1}{2} \right )
\right )$, $j=0,...,s-1$ and $R = \sum_{j=0}^{s-1}r_j$, satisfies
\begin{equation} \label{Greensdeltaeqn}
\left ( 1 + (-\partial^2)^s \right ) g^{(s)}(x) = \frac{2 s}{R} \delta (x)\,.
\end{equation}
The Green's function $g^{s}(x)$ has a discontinuity in its s-th
derivative.  The pulse shape (\ref{Greenseqn}) is plotted for $s=1$ peakons
and $s=2,3$ pulsons in Figure \ref{s123_pulsons_fig}.

\begin{figure}[ht!]
\centerline{
\scalebox{.4}{\includegraphics{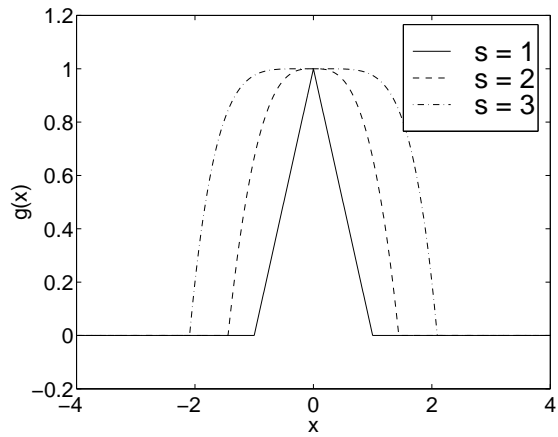}}
}
\caption{Compacton shapes for s=1,2, and 3.  These shapes widen as $s$
  increases.}
\label{s123_compactons_fig}
\end{figure}

In the high wave number limit, the Hamiltonian (\ref{Hseqn})
becomes
\begin{equation} \label{Hs_inf_eqn}
H_{s,k \rightarrow \infty} = \frac{1}{2}\int u^2_s\, dx \,,
\end{equation}
whose corresponding Green's function satisfies
\begin{equation} \label{Greens_inf_delta_neweqn}
(-\partial^2)^s g_c^{(s)}(x) = 2 s \delta (x)\,.
\end{equation}
The Green's functions corresponding to (\ref{Greens_inf_delta_neweqn}) having
compact support are given by
\begin{equation} \label{Greens_inf_eqn}
g_c^{(s)}(x) =
        \left\{ \begin{array}{cl}
        1 + \frac{s(-1)^s}{(2s-1)!} x^{2s-1} \hbox{sgn}(x) &
                \mbox{if
                $|x| < \left( \frac{(2s-1)!}{s}
                        \right)^{\frac{1}{2s-1}}$}\,, \\
        0 & \mbox{otherwise}.
        \end{array}
        \right.
\end{equation}
These are depicted in Figure \ref{s123_compactons_fig} according to
(\ref{Greens_inf_eqn}) for $s=1,3$, but the opposite convexity is
chosen for $s=2$. These
``compacton'' Green's functions serve as examples of traveling wave
solutions to (\ref{EPeqn}) with compact support.

In addition to the aformentioned Green's functions, we also test
a Gaussian and a composite Green's function
consisting of one central $s=2$ compacton and two outlying $s=1$
compactons.

\subsection{Numerical Method} \label{NMsection}

Equation (\ref{EPeqn}) is advanced in time with a fourth order Runge Kutta
scheme whose time step is chosen (by numerical trial and error) so as
to ensure less than a 1\% drop in
amplitude over 8 periodic domain traversals.  We keep track of the evolution
of the Fourier modes of $m(x,t)$ and compute the spatial derivatives in
(\ref{EPeqn}) pseudospectrally.  To transform between $u(x,t)$
and  $m(x,t)$ we convolve $m(x,t)$ with the Green's function in Fourier
space as $\hat{u}_n = N \hat{g}_n \hat{m}_n$, where $N$ is the
number of Fourier modes, which is kept at 2048 for each run. This
method of transforming between $m$ and $u$ proves convenient for arbitrary
Green's functions because the transformation can be computed numerically
and the relationship between $m$ and $u$ does not need to be determined
explicitly. Because antisymmetric perturbations to the zero solution are
unstable~\cite{CH[1993]},~\cite{CHH[1994]}, and the numerical
approximation of the nonlinear terms have aliasing errors in the high
wave numbers, we apply the following high pass filtered artificial
viscosity~\cite{Hyman}
\begin{equation} \label{HP_filter_eqn}
\nu(k) =
        \left\{ \begin{array}{cl}
        0 & \mbox{if $|k| < \frac{5N}{16}$} \\
        2\varepsilon(\frac{4 |k|}{N} - 1) &
             \mbox{if $\frac{5N}{16} < |k| < \frac{3N}{8}$} \\
        \varepsilon & \mbox{if $|k| > \frac{3N}{8}$}
        \end{array}
        \right. ,
\end{equation}
where $\varepsilon = 0.01$ for the present simulations.

\subsection{Initial Value Results} \label{iv_section}

Beginning with a normalized Gaussian distribution
$u(x,0) = \frac{1}{\sigma \sqrt{\pi} } e^{-{(x-x_0)^2}/{\sigma^2}}$
we first show the behavior of the initial value problem for the
integrable cases of peakons and $s=1$ compactons.
From~\cite{CH[1993]}\cite{CHH[1994]}, the asymptotic speeds of the emergent
integrable pulses can be determined as the eigenvalues of the Sturm-Liouville
problem
\begin{equation} \label{ISOeqn}
\psi_{xx}-\left[\frac{\delta_{CH}}{4}-\frac{1}{2\lambda}m(x,t)\right]\psi =
0,
\end{equation}
where $m(x,t)$ is the momentum density associated with the velocity
of the initial distribution and $\delta_{CH}$ is $1$ for the peakons
and $0$ for the $s=1$ compactons.  Camassa et al.~\cite{CHH[1994]} solve
(\ref{ISOeqn}) analytically with $m(x,0)=A\hbox{sech}^2(x)$.  The relation of
equation (\ref{ISOeqn}) to the spectral problem for the Schr\"odinger equation
in the case of positive ``potential,'' $m(x,t)$ has recently been discussed
in Beales et al.~\cite{BSZ[1998]}. See also Constantin~\cite{C[1998]} for a
similar discussion of the spectral problem for equation (\ref{ISOeqn}) in the
periodic case.

\begin{figure}[ht!]
\centerline{
\scalebox{1.0}{\includegraphics{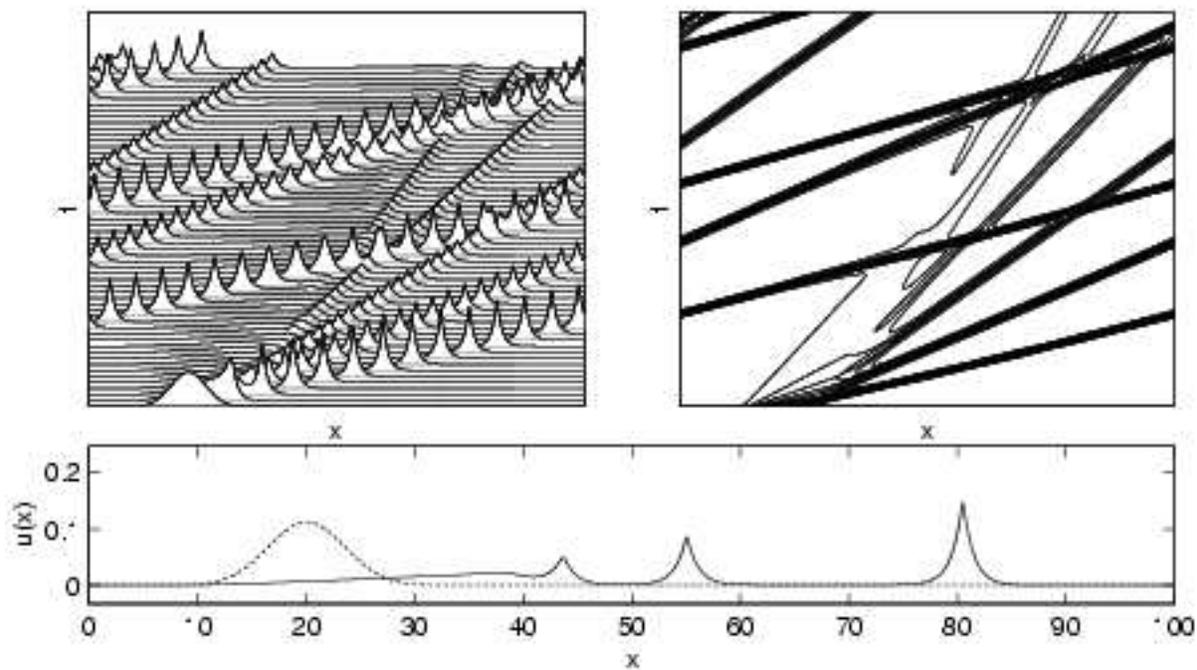}}
}
\caption{$s = 1$ peakons emerging from a Gaussian of unit area and
  $\sigma = 5$ centered about $x = 20$ on a periodic domain of
  length $L = 100$.  A secondary peakon emerges in the upper right hand corner
  of the contour plot as the faster peakons recross the domain and
  interact with the residual distribution.}
\label{iv_g_to_ps1_fig}
\end{figure}

\begin{figure}[ht!]
\centerline{
\scalebox{1.0}{\includegraphics{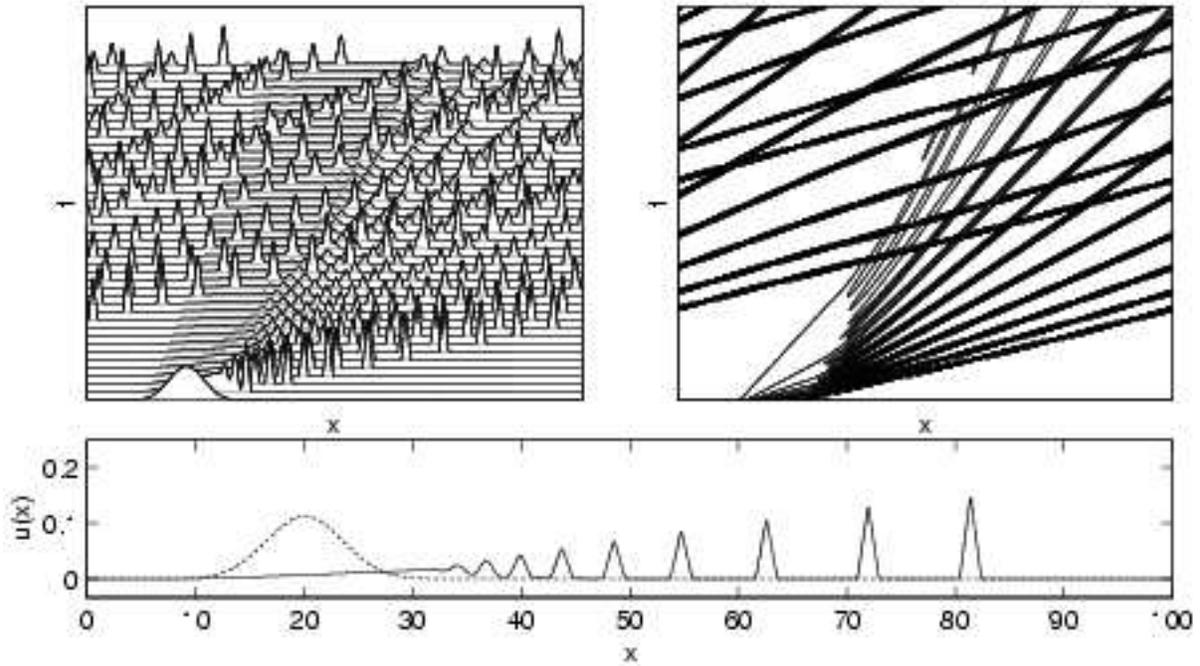}}
}
\caption{$s = 1$ compactons emerging from a Gaussian of unit area and
         $\sigma = 5$ centered about $x = 20$ on a periodic domain of
         length $L = 100$.  More compactons emerge here than peakons in
         Figure \ref{iv_g_to_ps1_fig}, because the compactons have
         smaller area.  The
         residual distribution around $x=20$ causes the
         space-time trajectories to curve as the larger, faster
         $s=1$ compactons interact with the slower ones that comprise the
residual.}
\label{iv_g_to_cs1_fig}
\end{figure}

In the simulations we investigate here the  asymptotic speeds are
given by the heights of the emergent pulses computed numerically from
(\ref{EPeqn}). Figures \ref{iv_g_to_ps1_fig} and
\ref{iv_g_to_cs1_fig} depict the emergence of peakons and $s=1$ compactons from
the initial Gaussian distribution.  The integrable behavior is evidenced as the
pulses in both cases collide elastically as they recross the periodic domain.
Note the presence of roughly nine distinct $s=1$ compactons in Figure
\ref{iv_g_to_cs1_fig} versus three peakons in Figure
\ref{iv_g_to_ps1_fig}.  While the maximum speed is identical for both cases,
the number of emergent pulses increases for pulses of smaller area.

In Figure \ref{iv_g_to_ps1_fig} we see the interaction of the larger
peakons with the slower residual distribution that is left behind
by the initial distribution after some early evolution has occurred.
Notice that the last two peakons emerging in the upper right hand
corner of the contour plot experience a positive phase shift (to the right)
shortly after the sixth peakon emerges.  This is consistent with the increase
in amplitude of the peakon as it grows out of the initial distribution
and propagates away, enhanced by the interaction of the larger, faster
peakons with the much slower residual distribution, whose peakon content
has not
yet emerged. This rather complex interaction in which the sixth peakon
experiences a significant amount of growth as it traverses the upper
right hand corner of the figure could
have been avoided by removing the residual distribution remaining after
the first few peakons had emerged. However, the Gaussian initial
distribution for this figure is comprised of a countable infinity of
peakons interacting nonlinearly among themselves.  Thus, removing the
residual distribution this way would alter the intended initial value
problem perhaps unacceptably.
Should we desire only a certain finite number of peakons, say, six of them, to
emerge from an initial distribution, then we could use
the inverse problem to determine what initial distribution is formed from
this number of pulses.  We discuss the inverse problem as well as the
reversible nature of equation (\ref{EPeqn}) later in the paper.

\begin{figure}[ht!]
\centerline{
\scalebox{1.0}{\includegraphics{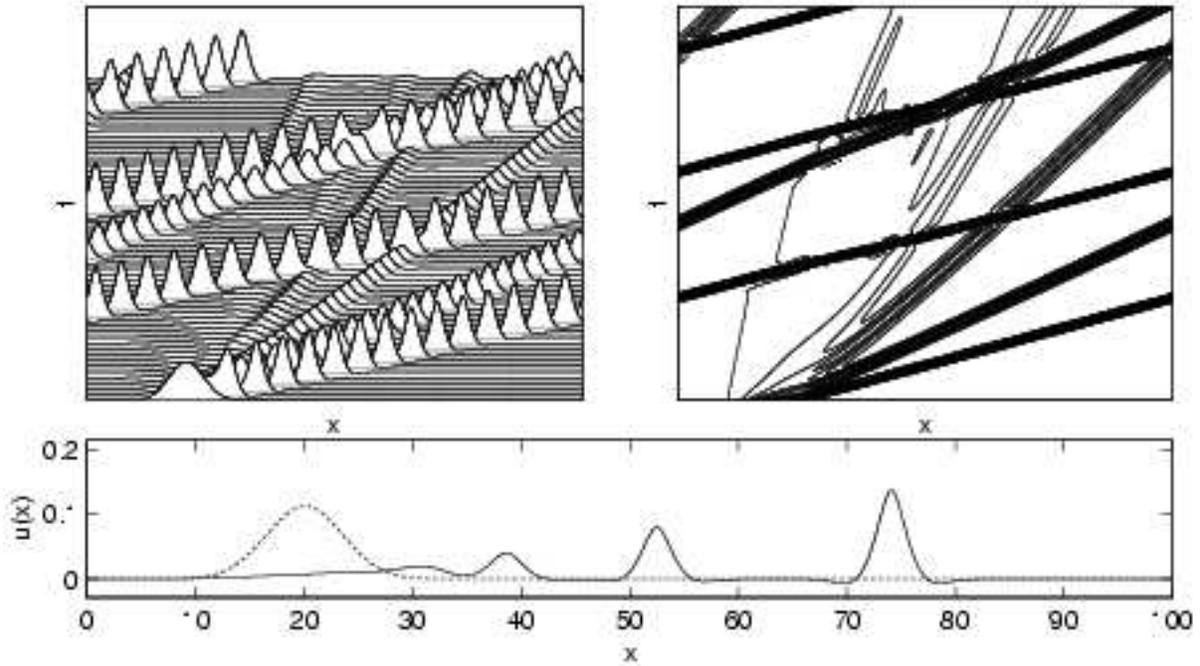}}
}
\caption{$s = 2$ pulsons emerging from a Gaussian of unit area and
         $\sigma = 5$ centered about $x = 20$ on a periodic domain of
         length $L = 100$.  The space-time trajectories
         curve just after the $s=2$ pulsons emerge from the initial
distribution
         because their large width enhances their interaction
         range.  Note the phase shift in the space-time plot occurring at
the trailing edge of the
         residual distribution as the larger, faster pulsons
         recross the domain.}
\label{iv_g_to_ps2_fig}
\end{figure}

\begin{figure}[ht!]
\centerline{
\scalebox{1.0}{\includegraphics{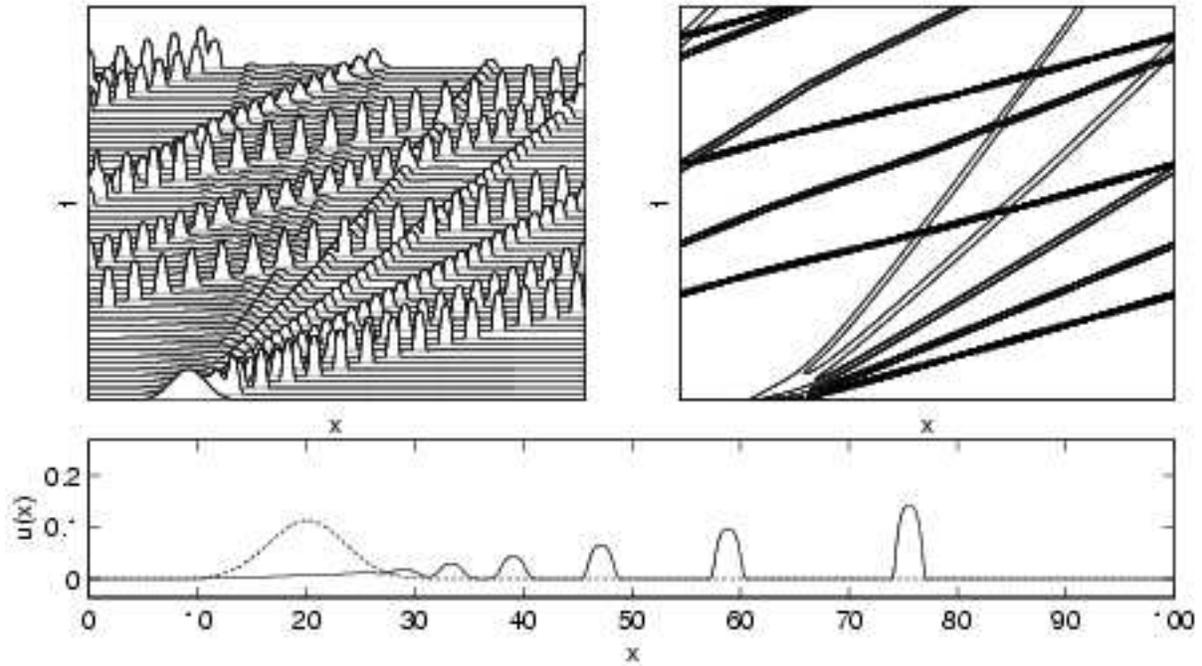}}
}
\caption{$s = 2$ compactons emerging from a Gaussian of unit area and
         $\sigma = 5$ centered about $x = 20$ on a periodic domain of
         length $L = 100$.  The space-time trajectories
         curve slightly as they emerge from the initial distribution
         and then straighten out when the compactons are no longer
         interacting.  The space-time trajectories curve again as
         the compactons recross the domain and interact with the
         residual distribution.}
\label{iv_g_to_cs2_fig}
\end{figure}

\begin{figure}[ht!]
\centerline{
\scalebox{1.0}{\includegraphics{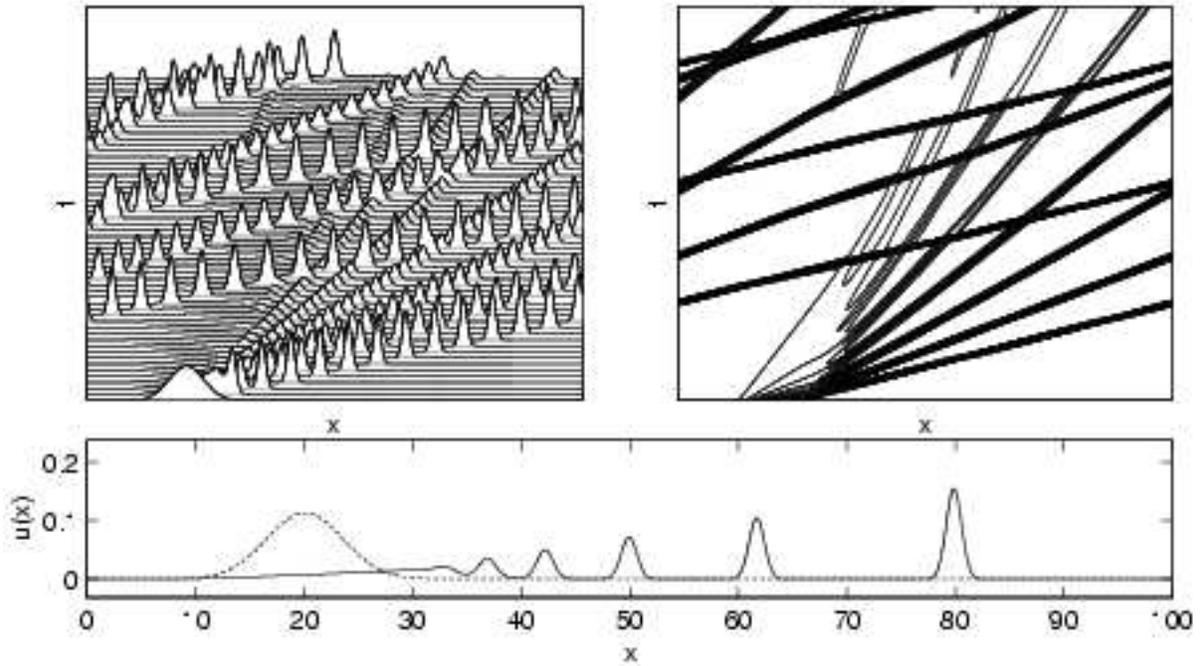}}
}
\caption{Gaussons emerging from an initial Gaussian of unit area and
         $\sigma = 5$ centered about $x = 20$ on a periodic domain of
         length $L = 100$. The interaction of the faster
         Gaussons with the residual ``ramp'' distribution causes another
Gausson
         to emerge at the top center of the contour plot.  The
         interaction between nearby Gaussons and between Gaussons and the
residual
         distribution causes their space-time trajectories to curve.}
\label{iv_g_to_g_fig}
\end{figure}

\begin{figure}[ht!]
\centerline{
\scalebox{1.0}{\includegraphics{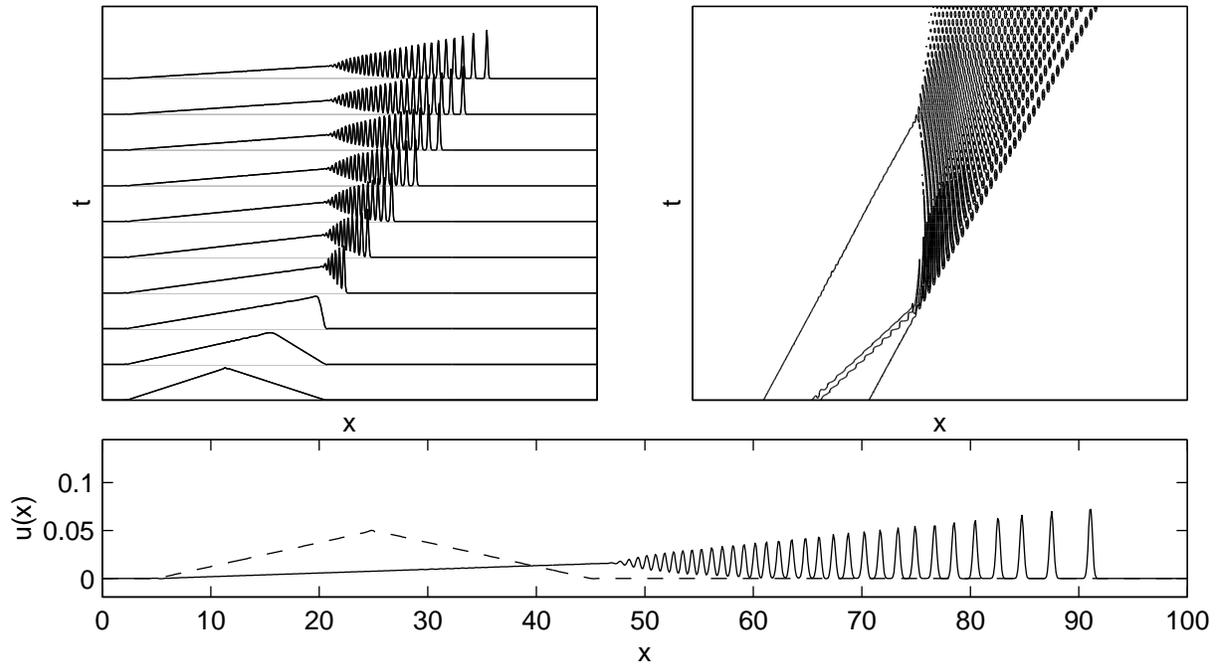}}
}
\caption{Gaussons emerging from an initial $s = 1$ compacton of unit area and
         width $40$ centered about $x = 20$ on a periodic domain of
         length $L = 100$.  In an effort to avoid
         confusion, the Gaussons are not shown recrossing the domain.}
\label{iv_cs1_to_g_fig}
\end{figure}

Next, we discuss the nonintegrable results in Figures \ref{iv_g_to_ps2_fig},
\ref{iv_g_to_cs2_fig}, \ref{iv_g_to_g_fig} and \ref{iv_cs1_to_g_fig}.
In Figure \ref{iv_g_to_ps2_fig} we see $s=2$ pulsons emerging from
an initial Gaussian velocity distribution.  Note the curvature of the
space-time trajectories in the contour plot for the third and fourth
pulsons.  This occurs because the
larger width of the $s=2$ pulsons compared to the peakons in Figure
\ref{iv_g_to_ps1_fig} implies a longer interaction period between
pulson pairs. Again we see the effects of the interaction of the larger
pulsons with the
residual distribution as they recross the periodic domain.  Most notably
we see the continued emergence of pulsons on the ``upwind'' side of the initial
distribution, leading to the emergence of a pulson at the top center of the
contour plot in Figure \ref{iv_g_to_ps2_fig}.  Figures
\ref{iv_g_to_cs2_fig} and \ref{iv_g_to_g_fig} also depict this
phenomenon of curvature in the space-time trajectories during the
emergence of the $s=2$ compactons and Gaussons.  In
Figure \ref{iv_g_to_cs2_fig} the curvature in the space-time
trajectories is not only because of the prolonged interaction between the
$s=2$ compactons,  but also because of their interaction with the residual
initial distribution as they recross the domain.  We do not show the Gaussons
emerging from the initial $s=1$ compacton shape and colliding with the initial
residual distribution in Figure \ref{iv_cs1_to_g_fig} because there are too
many pulses to make a clear plot.  However, this figure helps to support our
main conclusion: that an arbitrary (even) wave form given by the Green's
function $g(x)$ can be made to emerge from an (essentially) arbitrary confined
initial condition.  In this case we have chosen a wide $s=1$ compacton for the
initial distribution partly because its shallow discontinuity imposes less
stringent conditions on the numerics, and also to demonstrate the effects that
the width of the initial condition has on the number of emergent pulses.
Namely, the width of the initial distribution must be greater than the single
pulson width prescribed by the function $g(x)$, and wider initial distributions
produce a larger number of pulsons.

In summary, Figures \ref{iv_g_to_ps2_fig}, \ref{iv_g_to_cs2_fig},
\ref{iv_g_to_g_fig} and \ref{iv_cs1_to_g_fig} demonstrate that: (1) the shape
of each wave form emerging in the initial value problem is controlled by the
choice of $g(x)$; and (2) the dynamical behavior of the emergent pulses is
qualitatively the same as the integrable behavior of the peakons and $s=1$
compactons.

\subsection{Rear-End Collisions}

\begin{figure}[ht!]
\centerline{
\scalebox{1.0}{\includegraphics{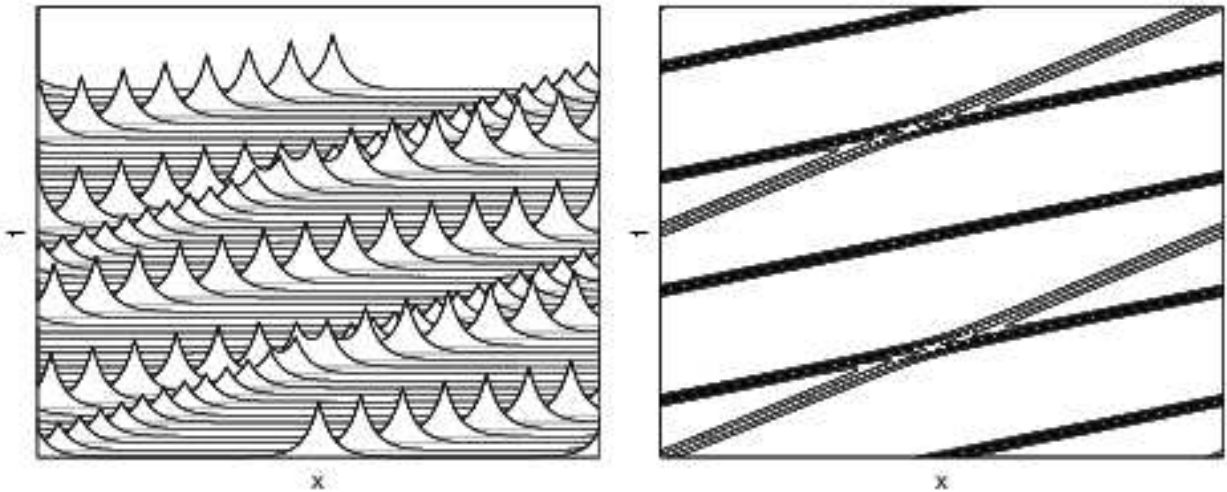}}
}
\caption{Rear-end collision dynamics for $s=1$ peakons.  The faster peakon
moves
  at twice the speed of the slower one.  For this case,
  both collisions result in a phase shift to the right for the faster
  space-time trajectory, but no phase shift for the slower one.  The phase
shift
  causes the second collision to occur slightly to the left of the first.}
\label{rear_puls_s1_fig}
\end{figure}

\begin{figure}[ht!]
\centerline{
\scalebox{1.0}{\includegraphics{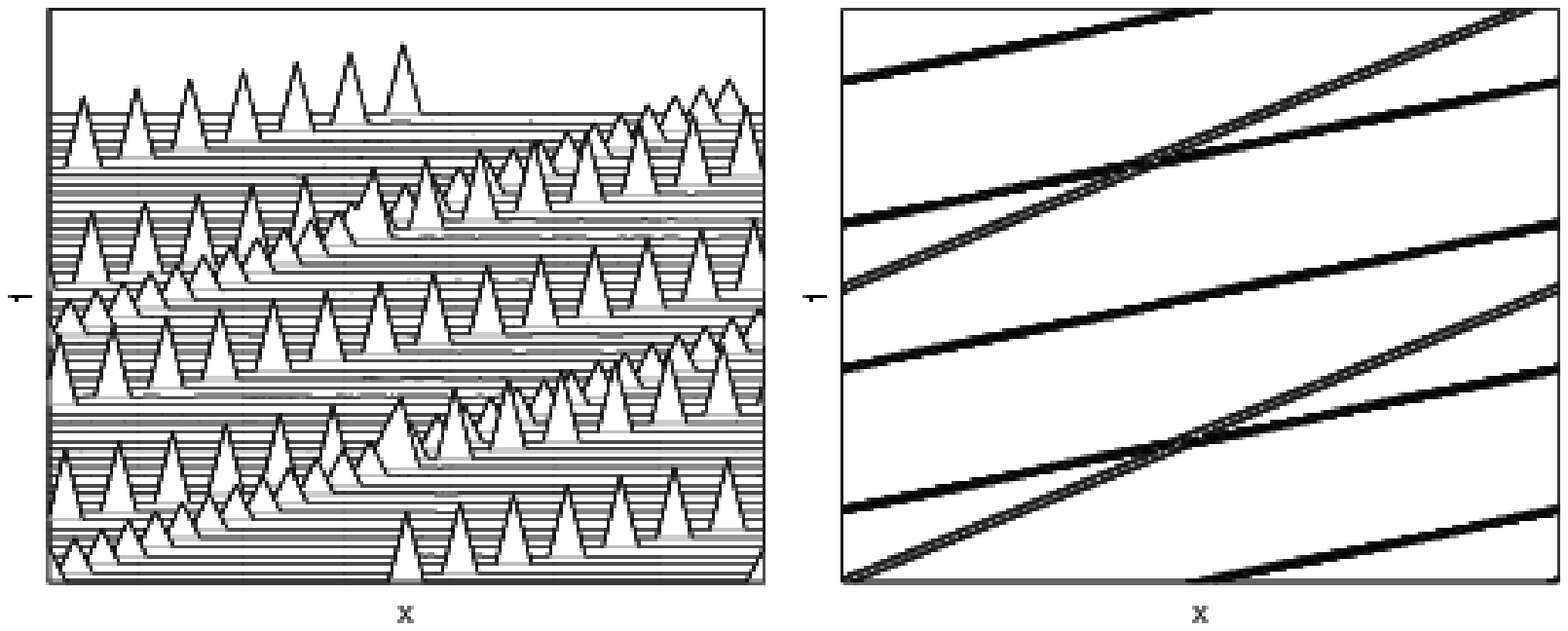}}
}
\caption{Rear-end collision dynamics for $s = 1$ compactons.
  The faster space-time trajectory experiences a phase shift to the
  right while the slower one experiences a phase shift to the left.}
\label{rear_comp_s1_fig}
\end{figure}

We compute the evolution governed by equation (\ref{EPeqn}) with two wave forms
(for a particular choice of Green's function) located initially at $x = 0$ and
$x = L/2$ with amplitudes $c_1=1$ and $c_2=1/2$, respectively. This situation
is arranged to investigate the nature of rear-end collisions both for the
integrable and nonintegrable cases.  Beginning with the integrable cases
depicted in Figures \ref{rear_puls_s1_fig} and
\ref{rear_comp_s1_fig}, we see that, consistent with the findings
of~\cite{CHH[1994]}, the slower peakon in Figure \ref{rear_puls_s1_fig}
experiences no phase shift, but the faster peakon experiences a shift
to the right.  This occurs after each collision, thus causing the second
collision to occur to the left of the first.  In Figure
\ref{rear_comp_s1_fig} we see that the faster $s=1$ compacton
trajectory experiences a phase shift to the right while the slower one
experiences a shift to the left. Thus, perhaps not unexpectedly, the phase
shift
of the interaction depends on the pulse shape.  The sudden leftward phase
shift of the slower trajectory indicates that momentum is being
transferred rightward rather quickly in the collision process across the entire
width of both wave forms.  This is something like the sudden transfer
of momentum in the collision of billiard balls.  For this reason, the
magnitude of the phase shift increases with pulse width.  Note that
the pulsons exchange momentum in each collision, rather than passing
through each other.

\begin{figure}[ht!]
\centerline{
\scalebox{1.0}{\includegraphics{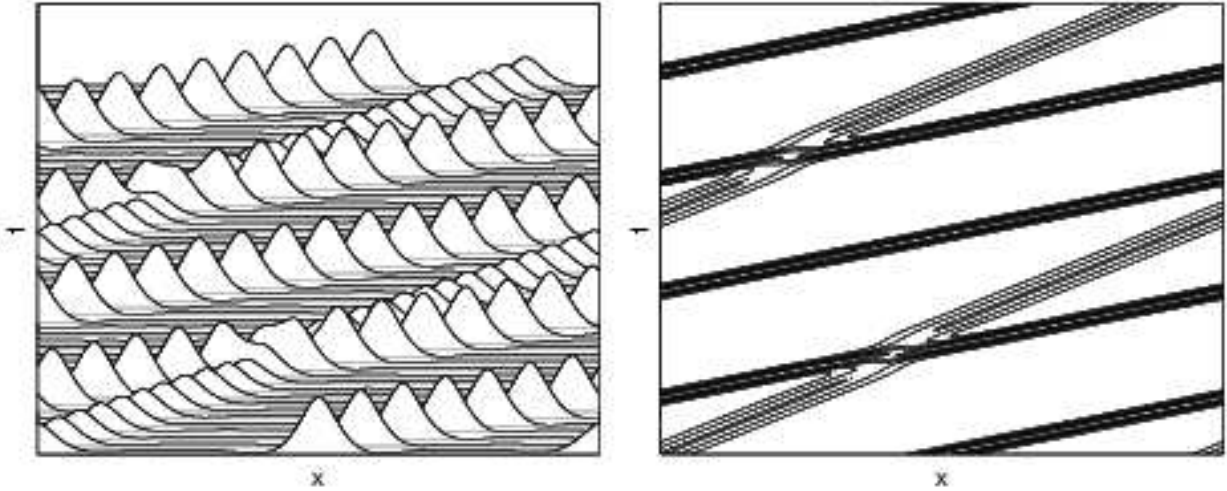}}
}
\caption{Rear-end collision dynamics for $s = 2$ pulsons.
  The faster space-time trajectory experiences a phase shift to the
  right while the slower one experiences a phase shift to the left.
  The size of the interaction region is proportional to the sum of the pulse
  widths.}
\label{rear_puls_s2_fig}
\end{figure}

\begin{figure}[ht!]
\centerline{
\scalebox{1.0}{\includegraphics{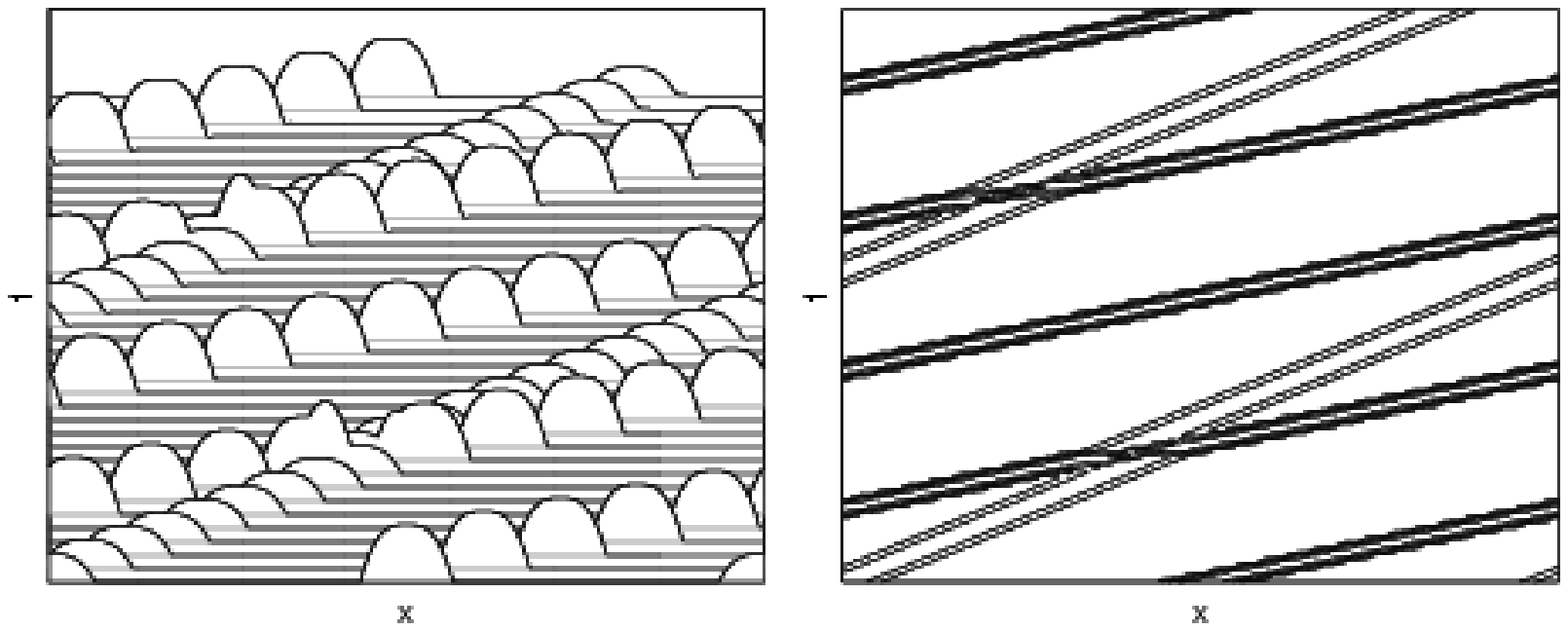}}
}
\caption{Rear-end collision dynamics for $s = 2$ compactons.
  The faster space-time trajectory experiences a phase shift to the
  right while the slower one experiences a phase shift to the left.}
\label{rear_comp_s2_fig}
\end{figure}

\begin{figure}[ht!]
\centerline{
\scalebox{1.0}{\includegraphics{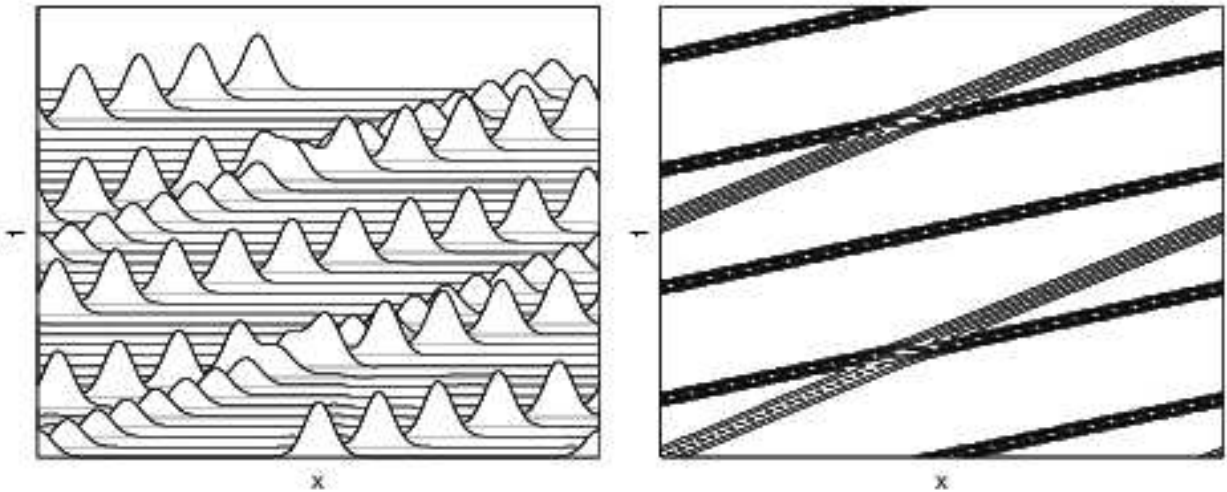}}
}
\caption{Rear-end collision dynamics for Gaussons.
  The faster space-time trajectory experiences a phase shift to the
  right while the slower one experiences a phase shift to the left.
  Because they are so narrow, Gaussons have the smallest phase shift
  of the four nonintegrable cases depicted in this figure and Figures
  \ref{rear_puls_s2_fig}, \ref{rear_comp_s2_fig}, and \ref{rear_multi_fig}.}
\label{rear_gaus_fig}
\end{figure}

\begin{figure}[ht!]
\centerline{
\scalebox{1.0}{\includegraphics{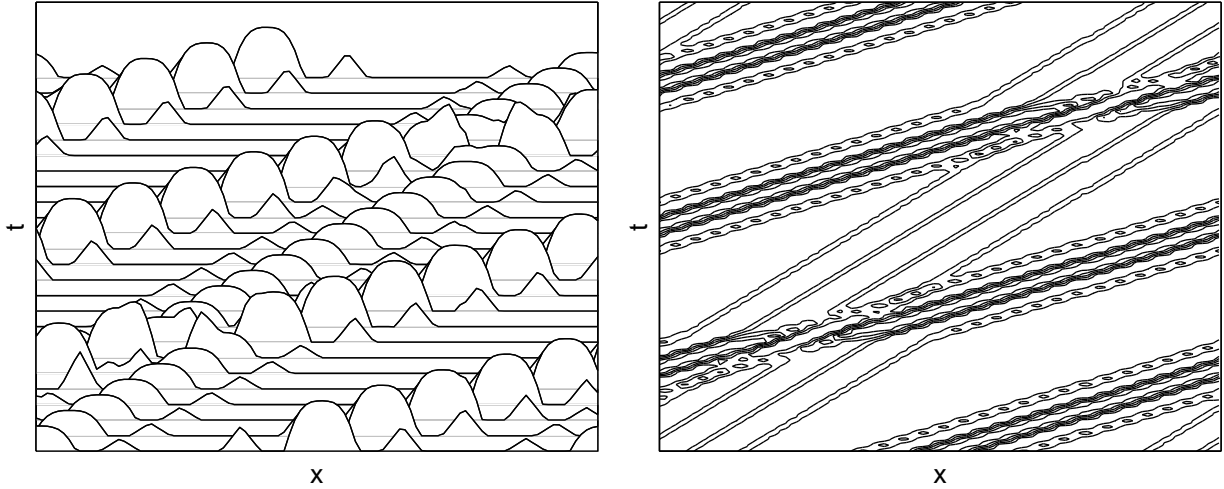}}
}
\caption{Rear-end collision dynamics for multicompactons.
  The faster space-time trajectory experiences a phase shift to the
  right while the slower one experiences a phase shift to the left.
  The large width and hence large interaction range of the multicompactons
  causes them to have the largest phase shift of the four
  nonintegrable cases depicted in this figure and Figures
  \ref{rear_puls_s2_fig}, \ref{rear_comp_s2_fig}, and \ref{rear_gaus_fig}.}
\label{rear_multi_fig}
\end{figure}

The nonintegrable cases depicted in Figures \ref{rear_puls_s2_fig},
\ref{rear_comp_s2_fig}, \ref{rear_gaus_fig}, and \ref{rear_multi_fig}
all show similar elastic collisions resulting only in phase shifts.
In each case the faster pulse trajectory experiences no shift, or a shift
to the right,
while the slower one experiences a shift to the left.  The magnitude of the
shift increases with pulse width.  The pulse width is the largest
for the multicompactons in Figure
\ref{rear_multi_fig} and they experience the largest phase shift of the four
nonintegrable cases shown here.  The Gaussons in Figure \ref{rear_gaus_fig},
on the other hand, experience the least phase shift and they are the
narrowest of the four nonintegrable pulsons we treat.

\subsection{Head-on Collisions} \label{HOsection}

\begin{figure}[ht!]
\centerline{
\scalebox{1.0}{\includegraphics{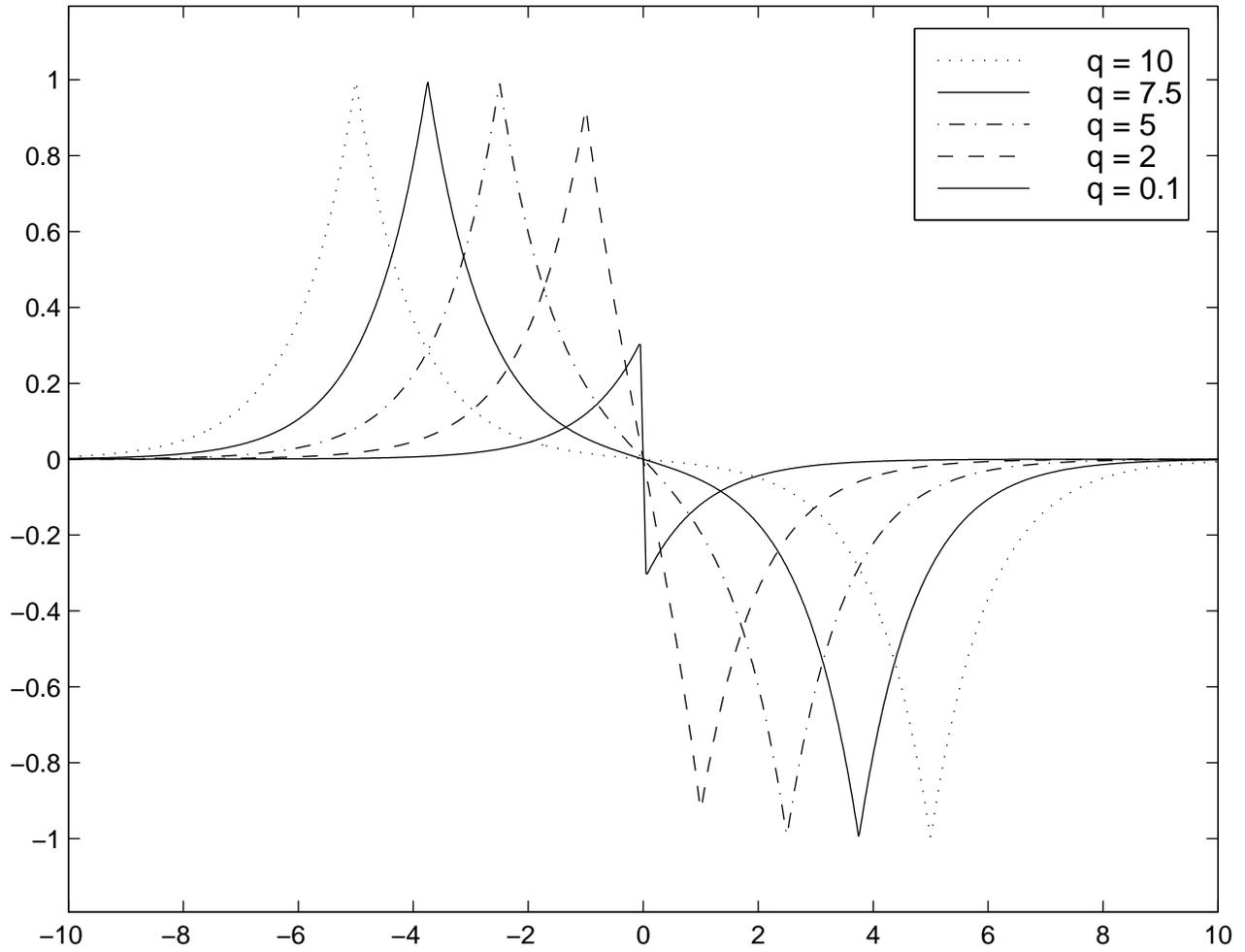}}
}
\caption{Head-on collision dynamics for $s=1$ peakons at five
  pre-collision separations $q$.  The velocity
  tends to zero as the separation between the peakons tends to zero
  and a verticality in slope develops at $x=0$, as $q\to 0^+$.  At
  this collision point, the solution reverses its polarity by flipping
  across the horizontal axis (so that $u\to -u$) and the peakons move
  apart in opposite directions.}
\label{col_puls_s1_fig}
\end{figure}

\begin{figure}[ht!]
\centerline{
\scalebox{1.0}{\includegraphics{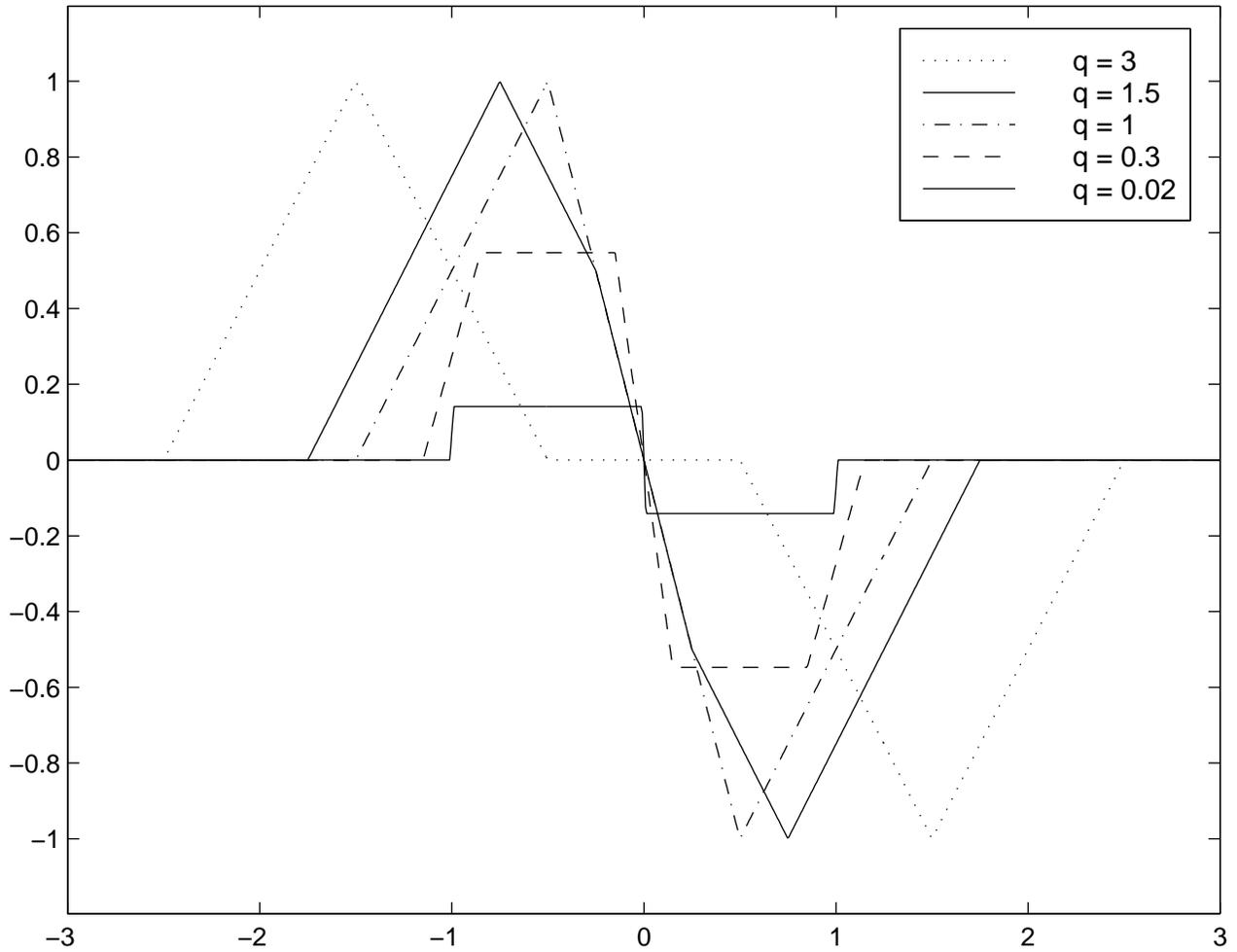}}
}
\caption{Head-on collision dynamics for $s=1$ compactons at five
  pre-collision separations $q$.  As the
  compactons meet, the slope at $x=0$ remains at $u_x(0,q)=-2$ until
  their peaks are clipped when $q=1$.  Thereafter the slope at $x=0$ begins
to increase
  and eventually becomes vertical when $q=0$ just as the velocity vanishes.  At
  this moment the solution reverses polarity and the
  compactons re-emerge, moving apart in opposite directions.}
\label{col_comp_s1_fig}
\end{figure}

We use the exact solution $u(x,q)$ for arbitrary Green's functions in equation
(\ref{vel_collide_eqn}) to determine the velocity profiles in
head-on antisymmetric pulson-antipulson collisions. The complexity of
the wave shapes and the strengths of the various singularities which
form according to equation (\ref{vel_collide_eqn}) during these
interactions for many choices of pulson shapes are beyond the
capabilities of most numerics.  For example, the spatial derivative of equation
(\ref{vel_collide_eqn}) at $x=0$ gives
\begin{equation} \label{ux_x0eqn}
u_x(0,q) = \frac{2c}{\sqrt{1-g(q)}}g'(q/2)\,.
\end{equation}
Hence, whether a vertical slope in $u$ forms at $x=0$ depends on
the choice of the Green's function $g(x)$.  Colliding  peakon-antipeakon
profiles
are depicted in Figure \ref{col_puls_s1_fig}.  As discussed
in~\cite{CHH[1994]} the solution
$u$ in this case tends to zero as $q \rightarrow 0$, as the slope
$\lim_{q\to0}u_x(0,q)$ diverges to minus infinity.  Also, $p\rightarrow
  \infty$ as $q\rightarrow 0$, so as to maintain constant energy and zero
total momentum in the particle system.  This divergence phenomenon
occurs similarly for the $s=1$ compactons, as shown in
Figure \ref{col_comp_s1_fig}.

As with the peakons, the colliding $s=1$
compactons develop a vertical slope in velocity at $x=0$ when $q~=~0$.  Upon
making initial contact when $q=2$, the two triangular pulses develop a
maximum negative slope at $x=0$ of  $u_x(0,q)\vert_{q\in[1,2]} = -2c$ which
remains  constant
until $q=1$, whereupon the two pulses begin to clip their peaks and
become trapezoidal as the slope at $x=0$ tends toward minus infinity
via $\lim_{q\to0^+} u_x(0,q) = -2c/\sqrt{|q|}$, cf. equation (\ref{ux_x0eqn}).
When $q = 0$, the two colliding pulses diminish to $u=0$ and then
``bounce'' apart to reverse the aformentioned process, so that
$u_x(0,q) = +2c/\sqrt{|q|}$ until $q=1$ with
$u_x(0,q)\vert_{q\in[1,2]} = +2c$.  Thereafter, the reformed pulses separate in
opposite directions.  Note that $q$ approaches zero, but does not change sign;
so the ``particles'' with phase space coordinates $p_i$, $q_i$,
$i=1,2$, keep their order, just as for the rear-end collisions.

\begin{figure}[ht!]
\centerline{
\scalebox{1.0}{\includegraphics{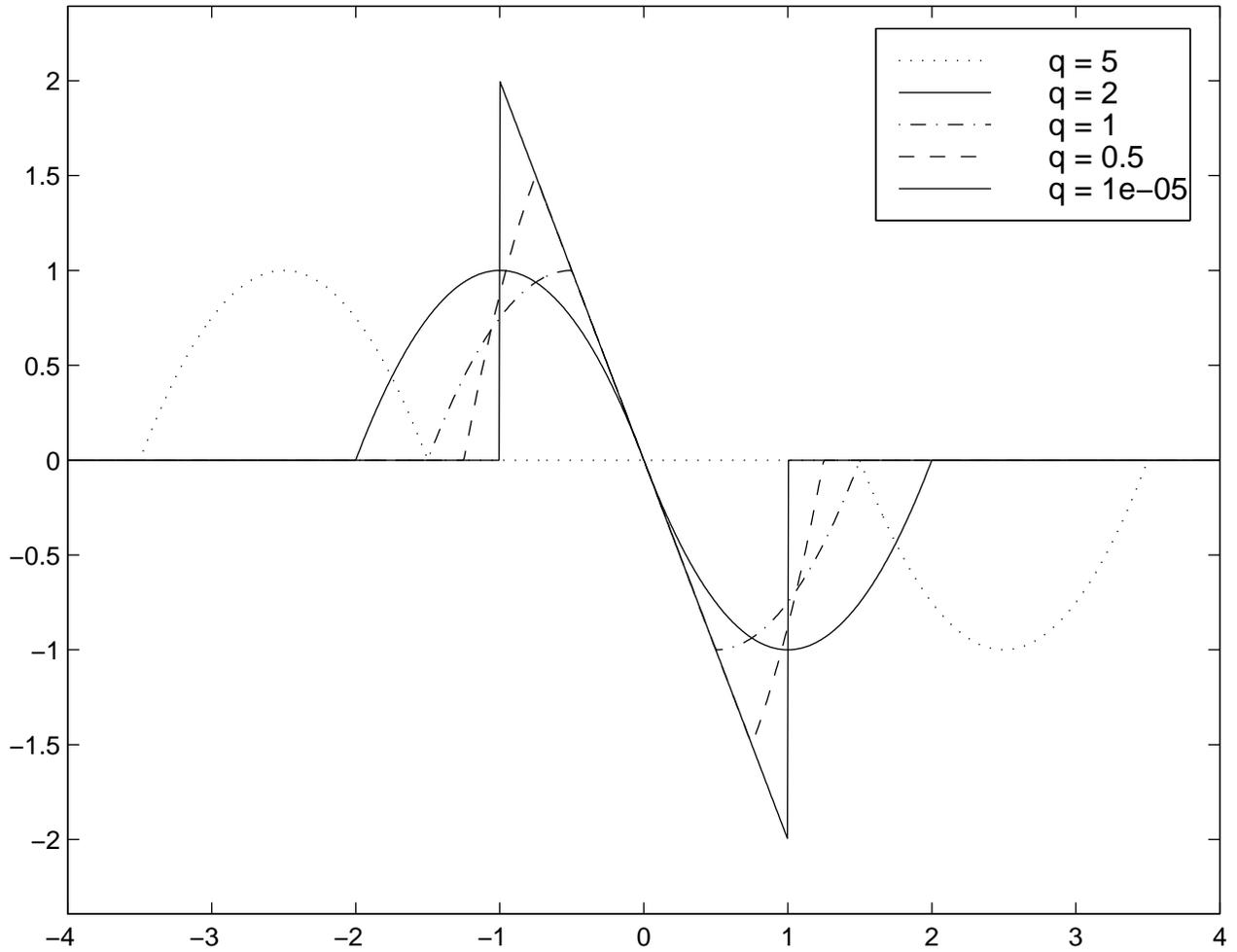}}
}
\caption{Head-on collision dynamics for parabolas at five
  pre-collision separations $q$.  Parabolas collide to form a finite
  distribution when $q=0^+$ and two verticalities in slope appear at $x=\pm
  1$.  After the collision the solution reverses polarity and the
  parabolas move apart, preserving the antisymmetry.}
\label{col_parab_fig}
\end{figure}

The most significant difference between the integrable and
nonintegrable pulson-antipulson collisions is that the solution $u(x,q)$
does not necessarily tend toward zero as $q \rightarrow 0$ for the
nonintegrable cases.  Another difference is that a verticality in
slope does not necessarily develop at $x=0$.  Moreover, such a
verticality may not develop at all.  As an example of a nonintegrable
pulse that does not approach $u=0$ as $q \to 0$, but does develop
{\it two} verticalities at that instant, we consider a parabolic pulse such
that
\begin{equation} \label{parab_eqn}
g(x) =
        \left\{ \begin{array}{cl}
        1-x^2 & \mbox{if $|x| \le 1\,,$} \\
        0 & \mbox{otherwise}.
        \end{array}
        \right.
\end{equation}
The antisymmetrically colliding parabolas are shown in Figure
\ref{col_parab_fig}.  As
the parabolic pulson collides with its antipulson, two verticalities
develop at $x~=~\pm~1$ rather
than at $x=0$.  By equation (\ref{vel_collide_eqn}), this interaction
creates a limiting
distribution such that
\begin{equation} \label{limit_parab_eqn}
\lim_{q \rightarrow 0^+} u(x,q) =
        \left\{ \begin{array}{cl}
        -2cx & \mbox{if $|x| \le 1\,,$} \\
        0 & \mbox{otherwise}.
        \end{array}
        \right.
\end{equation}

\begin{figure}[ht!]
\centerline{
\scalebox{1.0}{\includegraphics{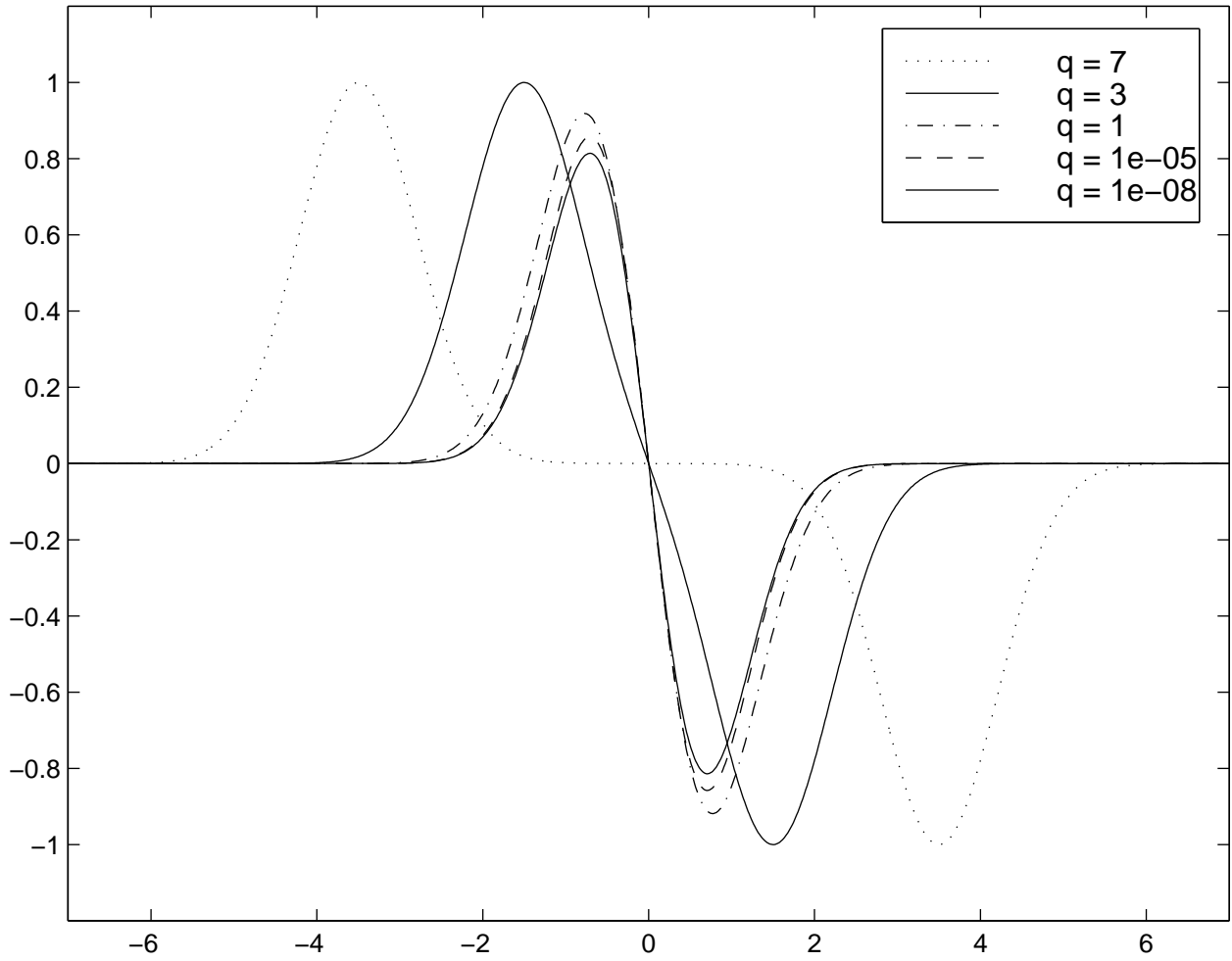}}
}
\caption{Head-on collision dynamics for Gaussons at five pre-collision
  separations $q$.  Gaussons collide to form a finite distribution
  when $q=0^+$ and develop no verticalities in slope.  After the
  collision the solution reverses polarity and the Gaussons move away from
each other.}
\label{col_gaus_fig}
\end{figure}

\begin{figure}[ht!]
\centerline{
\scalebox{1.0}{\includegraphics{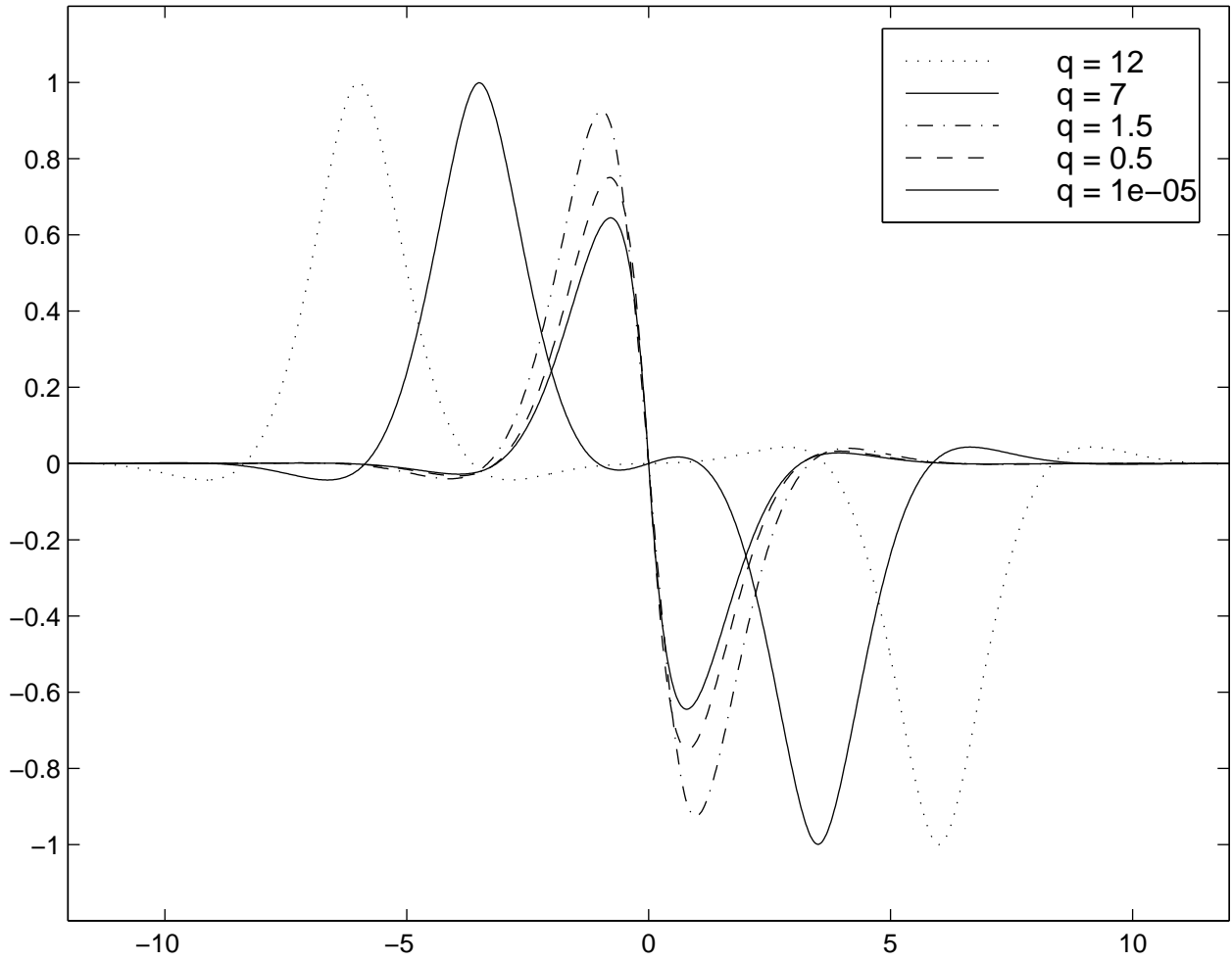}}
}
\caption{Head-on collision dynamics for $s=2$ pulsons at five pre-collision
  separations $q$.  As for the Gaussons in Figure \ref{col_gaus_fig},
  $s=2$ pulsons collide to form a finite distribution
  when $q=0^+$ and develop no verticalities in slope.  At the moment of
  collision, the solution reverses polarity and then the
  $s=2$ pulsons move apart antisymmetrically.}
\label{col_puls_s2_fig}
\end{figure}

As an example of a nonintegrable pulson that
does not tend toward $u=0$ and develops no verticalities in slope
as $q \rightarrow 0$, we consider the antisymmetrically colliding Gaussons
depicted in
Figure \ref{col_gaus_fig}.  Here the limiting distribution is
\begin{equation} \label{limit_gaus_eqn}
\lim_{q \rightarrow 0^+} u(x,q) = -2cx e^{-x^2}.
\end{equation}
When an antisymmetric collision results in a limiting distribution as
in equations (\ref{limit_parab_eqn}) and (\ref{limit_gaus_eqn}) as $q
\rightarrow 0^+$,
the solutions must flip instantaneously as they approach $q=0^+$, and
bounce apart so that the post-collision interaction will be identical but
opposite in
polarity to the pre-collision interaction.  This situation also occurs
similarly for the
antisymmetric collision of $s=2$ pulsons, shown in Figure
\ref{col_puls_s2_fig}.  However, the analytical solution for the
limiting $q \rightarrow 0^+$ distribution is not so easily expressed
in this case, so we omit it here.

\begin{figure}[ht!]
\centerline{
\scalebox{1.0}{\includegraphics{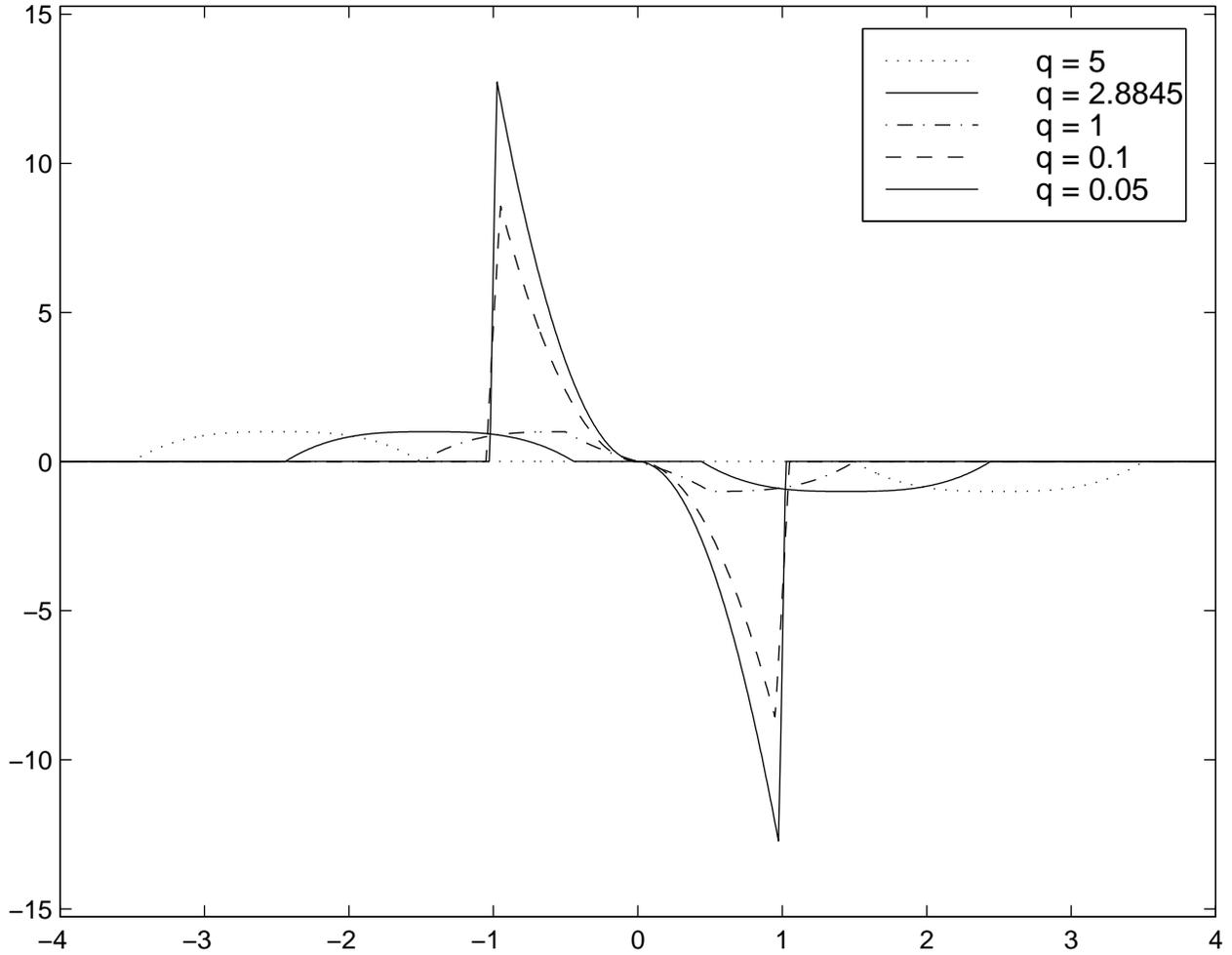}}
}
\caption{Head-on collision dynamics for $s=2$ compactons at five
  pre-collision separations $q$.  The $s=2$ compactons collide to form two
  verticalities in slope at $x = \pm 1$  as well as unbounded growth
  in $u(x,q)\vert_{x = \pm 1}$ as $q\to 0+$.  At this moment,
  the solution reverses polarity and the compactons move apart in opposite
directions.}
\label{col_comp_s2_fig}
\end{figure}

\begin{figure}[ht!]
\centerline{
\scalebox{1.0}{\includegraphics{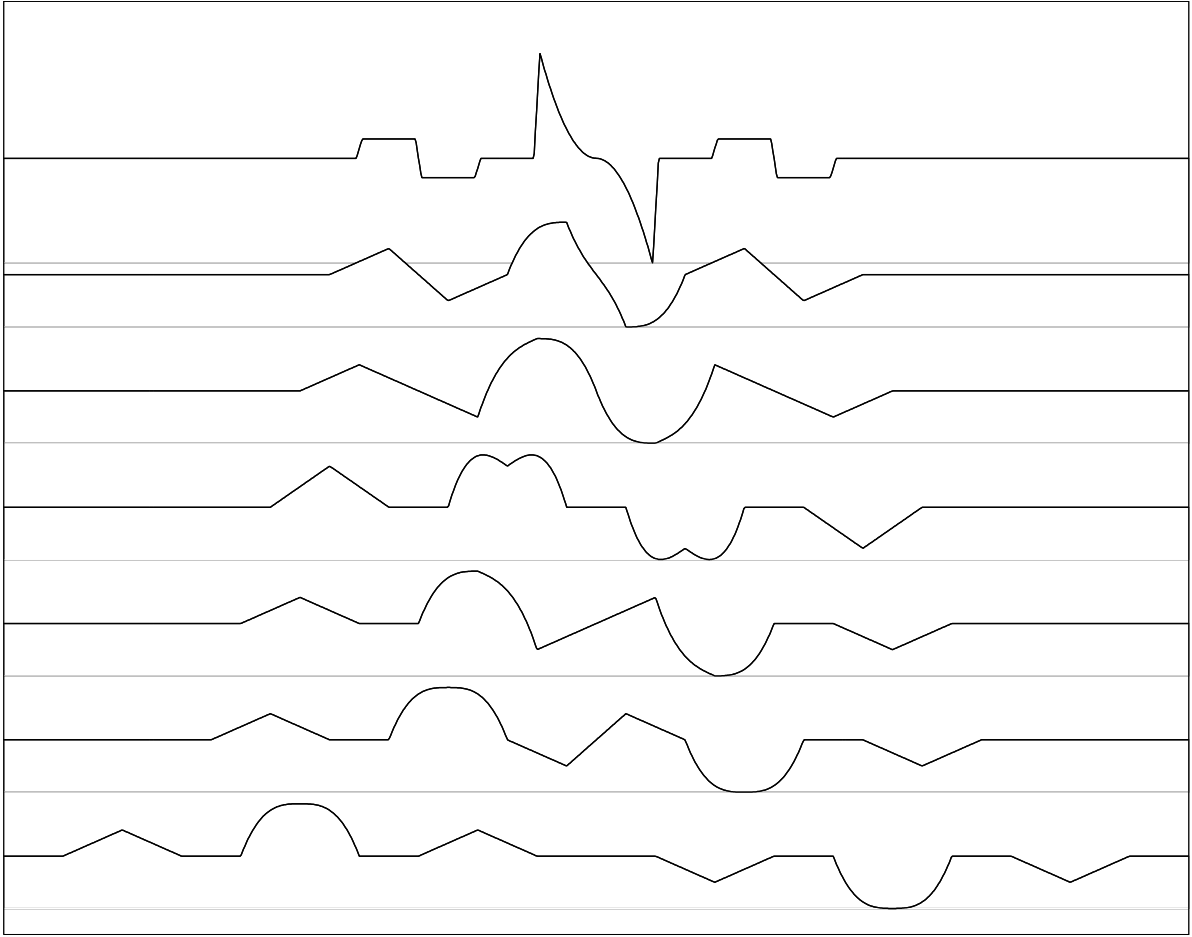}}
}
\caption{Head-on collision dynamics of multicompactons depicted at
separations of $q =$ 10, 5, 4,
  3, 2, 1, and 0.1.  We avoid confusion and show a ``waterfall'' plot to
depict the
  multicompactons rather than an overlay.  Here, verticalities as
  well as unbounded growth in $u(x,q)$ develop at $x=\pm 1$, $\pm
  2$, $\pm 3$, and $\pm 4$ as $q\to 0^+$.  The last distribution is
  scaled by a factor of $1/2$ to keep the large amplitudes resulting
  from the approaching unboundedness of $u(x,q)$ within the plot.}
\label{col_multi_fig}
\end{figure}

An even more complex situation develops when two $s=2$ compactons
collide antisymmetrically, as shown in Figure \ref{col_comp_s2_fig}.
In this case, not only do the colliding pulses develop two
verticalities, but there is no limiting distribution
as $q \rightarrow 0^+$.  In fact, one can show for $s=2$
compactons (with the factor of $1/3$ omitted for convenience) that
\begin{equation} \label{comp_s2_lim_eqn}
u(x = \pm 1,q) = \mp \frac{2}{\sqrt{q}}\left ( 1 + \frac{q^2}{8} \right ),
\end{equation}
so that the velocity at $x = \pm 1$ becomes unbounded as $q \rightarrow
0^+$.  However, as soon as $q$
reaches zero, the unbounded solution for $u$ switches sign and the pulses
restore themselves with reversed polarity and move apart.  As a final
example we show in Figure \ref{col_multi_fig} the
antisymmetric collision of ``multicompactons,'' again with the factor of
$1/3$ omitted from the central hump for convenience.  Here we use a
``waterfall''
plot to depict the collision since an overlay would be too confusing.  This
collision
results in the creation of eight verticalities at $x=\pm 1$,
$\pm 2$, $\pm 3$, and $\pm 4$.  Again, as $q\to 0^+$, the
solution flips polarity instantaneously and the interaction reverses as the
multicompactons move
apart.

\subsection{Reversibility} \label{RandS}

The results in Section \ref{iv_section} for the initial value problem
show that a train of pulses in the prescribed pulse shape $g(x)$
emerges from any initial distribution and dominates the
initial value problem.  In this process, the initial distribution breaks up
into a discrete number of pulsons, each of which travels at a speed
equal to its height, and the pulsons collide elastically among each other as
they recross the periodic domain.  An interesting feature of equation
(\ref{EPeqn}) is that it is time reversible, that is, invariant under $t\to -t$
and $u\to -u$. Thus, replacing $dt$ with $-dt$ at any
point in the numerical simulations causes the pulsons emerging in the
initial value problems to reverse their evolution sequence (but not their
heights) and collapse together into the initial distribution.  Thus the
time-reversed series  of collisions reforms the original initial distribution
at $t=0$. This process enables us in principle to determine the parameters
$p_i$, $q_i$, $i=1,\dots,N$, on the
invariant manifold by running the pde (\ref{EPeqn}) forward in time until
$N$ pulsons have emerged, then reversing the solution to recreate
the original smooth initial distribution.  Similarly, we may run the
pde forward until time $T$, determine the parameters ($p_i, q_i$) at
this time, remove any residual velocity distribution from which
pulsons have not yet emerged, then time-reverse the ordinary
differential equation (ode) dynamics for a period $T$ to create an
initial distribution containing only the pulson content that will
emerge before time $T$.  Should this process, denoted
in obvious notation as $[ODE(-T)\circ\Pi_N\circ PDE(T)]u(x,0)$, produce an
accurate approximation $u_N(x,0)$ of the original distribution $u(x,0)$, it
would be a useful procedure for approximately determining the pulson invariant
manifold content of a given initial velocity distribution.

\begin{figure}[ht!]
\centerline{
\scalebox{1.0}{\includegraphics{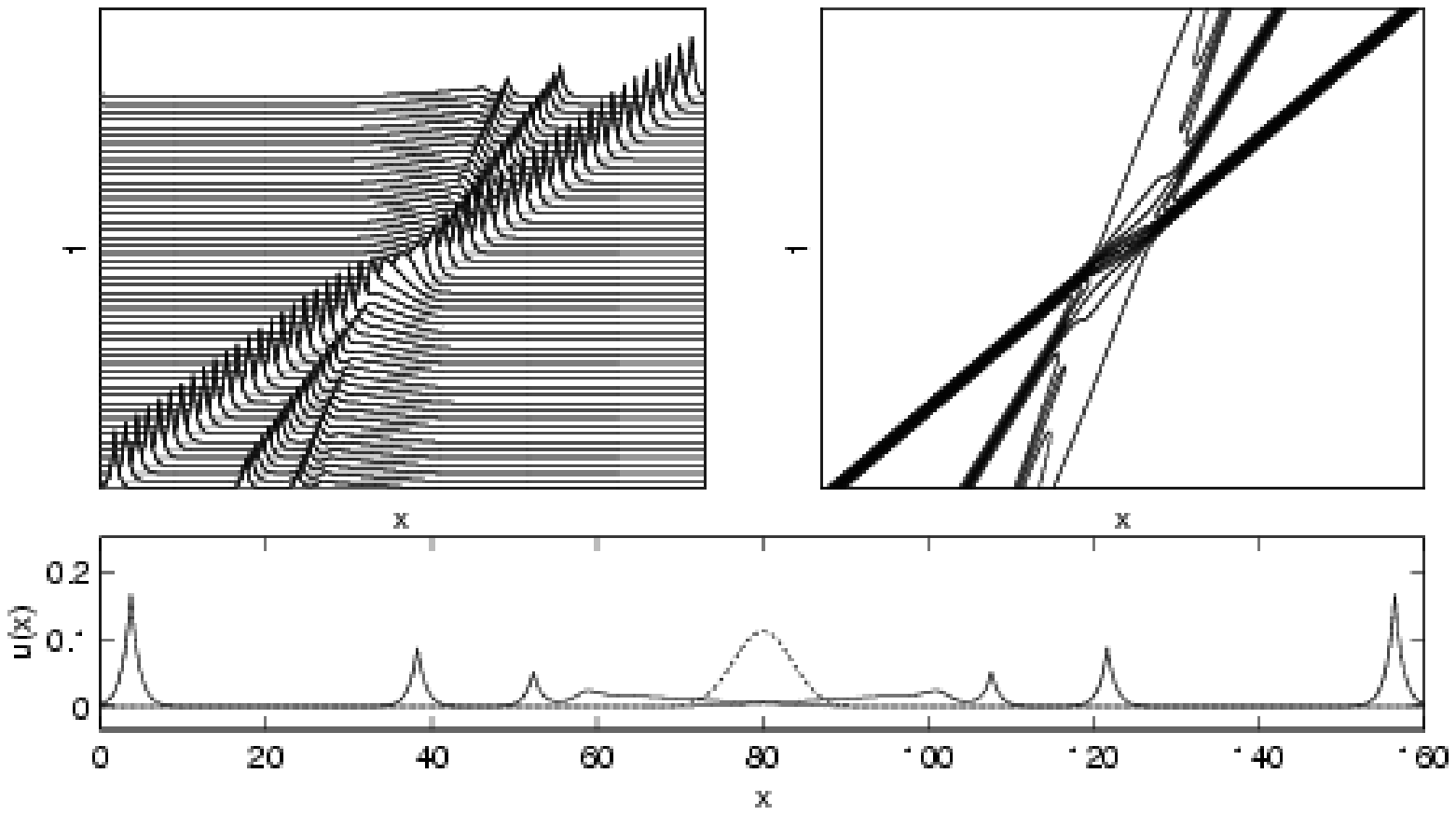}}
}
\caption{$s=1$ peakons colliding to form a Gaussian and then
  reemerging symmetrically thereafter.}
\label{iv_rev_ps1_fig}
\end{figure}

\begin{figure}[ht!]
\centerline{
\scalebox{1.0}{\includegraphics{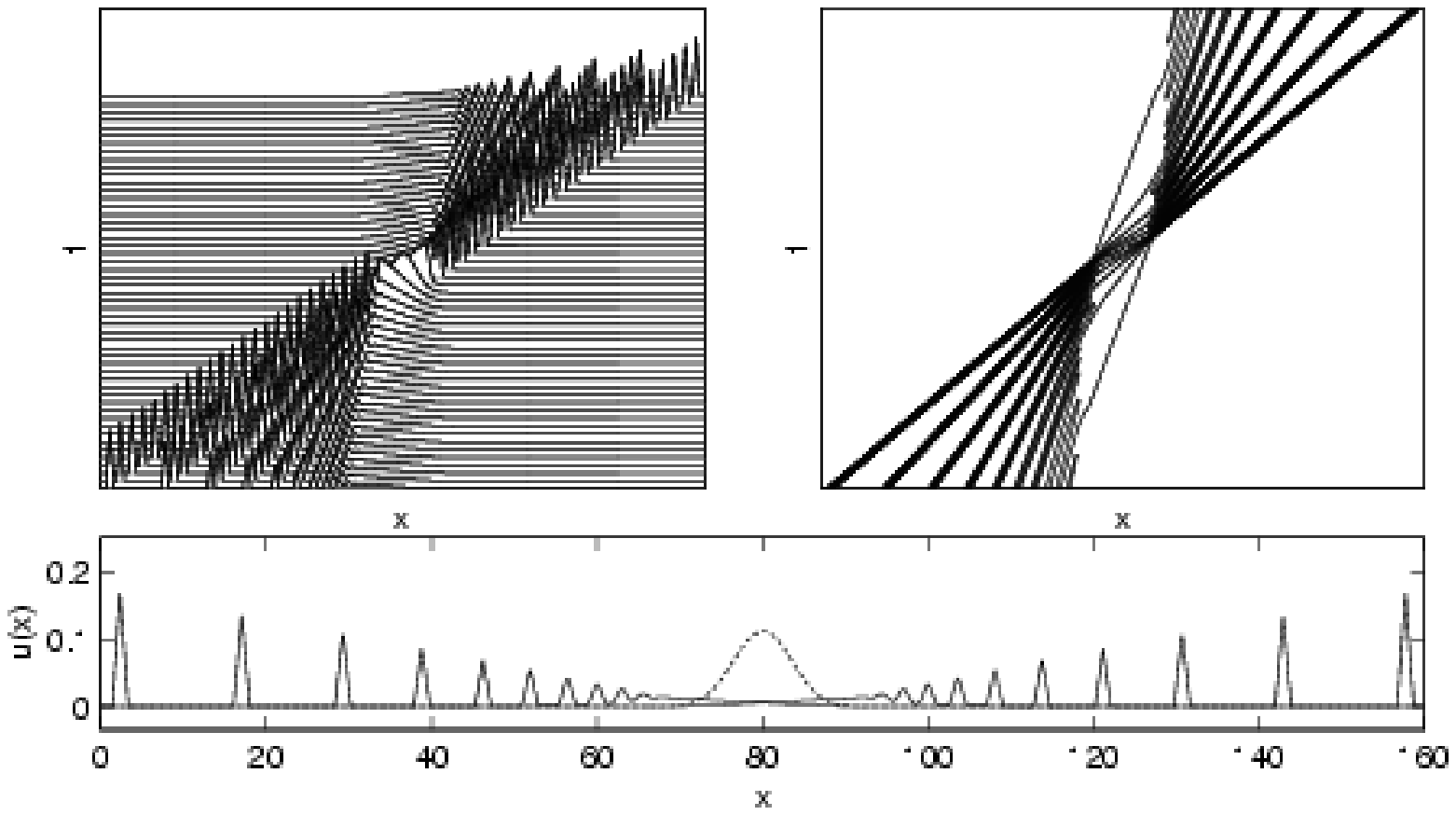}}
}
\caption{$s=1$ compactons colliding to form a Gaussian and then
  reemerging symmetrically thereafter.}
\label{iv_rev_cs1_fig}
\end{figure}

Figures \ref{iv_rev_ps1_fig}
and \ref{iv_rev_cs1_fig} were constructed by running the pde forward
in time from an initial Gaussian distribution until trains of three peakons
in Figure \ref{iv_rev_ps1_fig} and nine $s=1$ compactons
in Figure \ref{iv_rev_cs1_fig} emerged.  At this point time was reversed,
and the pde was run {\it in reverse} for twice as long, then the solution
was reflected in space. In this process trains of pulsons in reversed order
(with the tallest ones behind) reassembled moving rightward into the original
initial condition and then re-emerged as before (with the tallest ones
ahead).  The same ``geodesic pulson scattering'' figures could be generated
by running the initial Gaussian distribution both forward and backward
in time, then superposing the results.  Formally we have
$S(t) R_x S(t) = Id$ (where $S(t)$ is pde evolution for time $t$ and $R_x$ is
reflection in space), for initial conditions that are symmetric about the
origin.  Thus these figures could also have been generated by the process of
evolution, reflection, then evolution again.
This time reversal symmetry property indicates that the initial distribution
may be regarded not just as a sum of pulsons, but as a moment in time at which
a set of pulsons moving geodesically has collided to form a smooth
distribution,
in this case a Gaussian.

\section{Conclusions}

In equations (\ref{EPeqn}) and (\ref{Ueqn}) we have introduced a new family of
hyperbolic pde describing a certain type of geodesic motion and for some
members of this family we have studied its particle-like solutions. We call
these solutions ``pulsons.'' We have shown that, once initialized on their
invariant manifold (which may be finite-dimensional), the pulsons undergo
geodesic dynamics in terms of canonical Hamiltonian phase space variables. This
dynamics effectively reproduces the classical soliton behavior. We conjecture
that this behavior occurs because of the preponderance of two-body
elastic-collision interactions for the situation we consider of confined pulses
and confined initial conditions.

The dynamics we study in this framework of geodesic motion on a
finite-dimensional invariant manifold seems to account for all of the classical
soliton phenomena, including elastic scattering, dominance of the initial value
problem by confined pulses and asymptotic sorting according to height -- all
without requiring complete integrability. Thus, complete integrability
is apparently not necessary for exact soliton behavior. Regarding the
``formation'' of the pulsons: in fact, the pulsons must always be present,
since they compose an invariant  manifold. One discerns them in a given
(confined) initial condition by using the isospectral property in the
integrable cases, or just by waiting for them to emerge under the pde dynamics
in the nonintegrable cases. However, there is no ``pattern formation'' process
in this dynamics. Rather, there is an ``emergence of the pattern,'' which is
the pulson train.

This conclusion is illustrated by Figures \ref{iv_rev_ps1_fig} and
\ref{iv_rev_cs1_fig} showing geodesic scattering of an incoming set of
pulsons that collapses into a smooth ``initial'' distribution, then fans out
again into an outgoing set of pulsons that is the mirror image of the
incoming set. This parity-reflection is an implication of reversibility, as
well; since reversing the dynamics of a set of separated pulsons reassembles
them into a smooth ``initial'' distribution, which is just their sum, with
appropriate values for their initial ``moduli'' parameters
$p_i(0),q_i(0)$, $i=1,\dots,N$. The moduli parameters $p_i,q_i$ are collective
phase space coordinates on an invariant manifold for the pde motion. Once
initialized, these collective degrees of freedom persist and emerge as a train
of pulses, arranged in order of their heights. Moreover, since their
velocity is their peak height, small residual errors departing from the pulson
superposition in the assignment of initial parameter values $p_i(0),q_i(0)$,
$i=1,\dots,N$, do not propagate significantly. On the real line, such residual
errors are simply left behind by the larger pulsons traveling more quickly.
And on the periodic interval, these errors provide an occasional phase shift
perturbation to the stable pulsons.  These perturbations are seen in the
figures as regions of curvature in the space-time trajectories of the pulsons.

Our main conclusions regarding the elastic collisions of the pulsons and their
role in the pde initial value problem are the following:
\begin{enumerate}
\item Two-pulson interactions are elastic collisions that conserve total
kinetic
energy and transfer momentum, as measured by the asymptotic speeds and peak
heights of the pulsons. The motion of the pulsons is governed by geodesic
kinematics of particles with no internal degrees of freedom; but with an
interaction range, or spatial extent, which is given by their Green's function,
or wave form, $g(x)$. Thus, these interactions are analogous to the collisions
of billiard balls. This is especially clear for pulsons of compact support.

\item The exact solution in equation (\ref{vel_collide_eqn}) determines
analytically the complex wave shapes and the strengths of the various
singularities that may form during antisymmetric head-on collisions of pulsons
and antipulsons. These singularities are verticalities in slope at which the
velocity may diverge, take a finite value, or even vanish, depending on the
choice of Green's function. At the moment of collision, the pulson and
antipulson ``bounce'' apart and the polarity of the antisymmetric wave form
reverses, $u\to-u$. Again this is analogous to billiards, but with the
difference that the reversal of momentum reverses the polarity of the pulson
velocity profile. Note that the wave forms reverse polarity and bounce apart;
they do not actually ``pass through'' each other, although they may appear
to do so.

\item The dynamics on the pulson invariant manifold is dominated by the
preponderance of two-body collisons. To the extent that the initial value
problem for the pde takes place on the pulson invariant manifold, these
two-pulson collisions should also dominate the solution of the pde
dynamics.
We conjecture this is so for most choices of confined pulse and initial
conditions of finite extent for the family of pde (\ref{EPeqn}) with $u$ given
in (\ref{Ueqn}).
\end{enumerate}

Thus, geodesy governs the motion and explains the elastic-scattering
soliton phenomena we observe, without providing (or even requiring there
exists) a means of analytically solving the initial value problem for this
family of pde. Even the motion on the pulson manifold is not integrable for
$N>2$, except in the cases corresponding to the Camassa-Holm, and Dym
integrable pde. Similar geodesic but nonintegrable families of equations
showing soliton behavior on a finite-dimensional invariant manifold may exist
for other integrable geodesic equations, such as KdV.

\section{Acknowledgments} This work was supported in part by the 
Computational Science Graduate Fellowship Program of the Office of 
Scientific Computing in the Department of Energy.  For their kind 
attention, suggestions and encouragement, we are happy to thank 
R. Camassa, P. Constantin, P. Fast, C. Foias, I. Gabitov, J. M. Hyman, 
Y. Kevrekidis, S. Shkoller, P. Swart and S. Zoldi.   

\section{Appendix on Numerical Stability}

As discussed in Section \ref{NMsection}, we add dissipation in the
higher modes for the simulations covered in this paper because the
Fourier pseudospectral method used is unstable to perturbations at the
grid scale.  These perturbations arise as a result of the Gibbs
phenomenon which causes an overshoot to occur at points of discontinuous
slope in
$u$.  Such an overshoot forms negative values of the function $u$,
which can artificially create antipulsons that emerge and interact with
positive pulsons to form verticalities in slope as described in
Section \ref{HOsection}.  As an example, we tested the stability of
the numerical method by initializing the domain with a perturbation at a
single
grid point, as
\begin{equation}
u_i =
        \left\{ \begin{array}{cl}
        10^{-3} & \mbox{if $i = N/2\,,$} \\
        0 & \mbox{otherwise},
        \end{array}
        \right.
\end{equation}
for $i = 1,\dots, N$.  This initial distribution was unstable and
caused exponential growth in the neighboring grid points, such that
$u_{N/2-1} \to +\infty$ and $u_{N/2+1}\to -\infty$, resulting from the
creation of a pulson-antipulson pair whose peaks grew unstably.  We
then ran the same initial condition with a dissipation coefficient of
$\varepsilon = 0.01$ in equation (\ref{HP_filter_eqn}), which stabilized
it.  We also ran it again with a nonzero linear dispersion coefficient
of $\kappa = 0.01$ in equation (\ref{EP_kappa_eqn}) which stabilized it as
well.  Adding
dissipation damps the higher modes as specified in Section \ref{NMsection}
while
adding linear dispersion smooths discontinuities in $u$, thus
effectively damping the higher modes.  Thus, either dissipation or
dispersion will suppress instability due to perturbations at the grid scale.
\begin{figure}[ht!]
\centerline{
\scalebox{.5}{\includegraphics{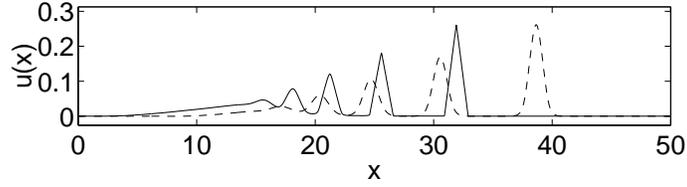}}
}
\caption{$s = 1$ compactons emerging from a Gaussian of width $\sigma
  = 3$ and unit area centered about $x = 10$ on a periodic domain of length $L
  = 50$ demonstrating the effects of the linear dispersion coefficient
  $\kappa$.  The solid line depicts the $\kappa = 0$ case with added
  dissipation while the dashed line depicts the $\kappa = 10^{-3}$
  case with no dissipation.}
\label{kappa_nonzero_fig}
\end{figure}

\begin{figure}[ht!]
\centerline{
\scalebox{.5}{\includegraphics{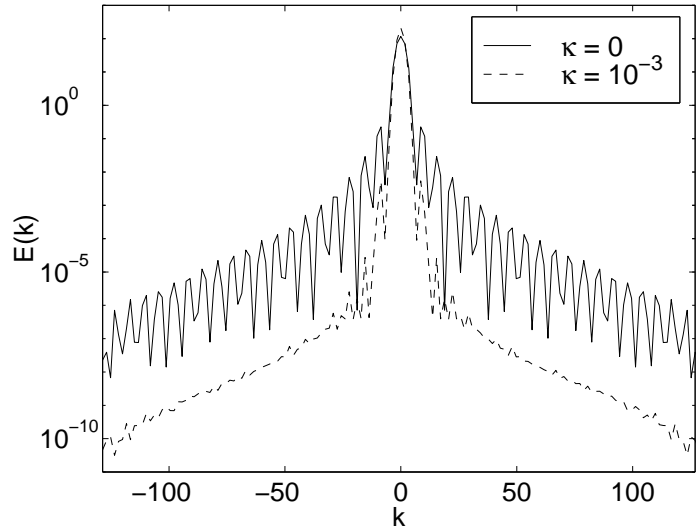}}
}
\caption{Power spectra of one pulse for the $\kappa = 0$ and $\kappa =
  10^{-3}$ cases.}
\label{kappa_E_k_fig}
\end{figure}

Figure \ref{kappa_nonzero_fig} compares dissipative and dispersive
dynamics for the case of $s~=~1$ compactons emerging from an initial
Gaussian velocity distribution.  Linear dispersion ($\kappa \ne 0$) acts to
smooth out discontinuities as well as hastening the emergence
time of each pulson.  This is demonstrated by the separation between
the leading pulses for the $\kappa = 10^{-3}$ and $\kappa = 0$ cases.
Despite the negligible difference in speed between the two,
the leading $\kappa = 10^{-3}$ pulson is far ahead of the
leading $\kappa = 0$ pulson, because the former was emitted sooner. Stability
is acquired by suppressing the higher modes, as shown in Figure
\ref{kappa_E_k_fig}.  Here the power spectrum of one compacton is shown for
both values of $\kappa$. Introduction of linear dispersion with $\kappa =
10^{-3}$ damps  the higher modes by roughly three orders of magnitude.

\clearpage

\section{Figures}

\clearpage

\clearpage

\clearpage

\clearpage

\clearpage

\clearpage

\clearpage

\clearpage

\clearpage

\clearpage

\clearpage

\clearpage

\clearpage

\clearpage

\clearpage

\clearpage

\clearpage

\clearpage

\clearpage

\clearpage 

\begin{thebibliography}{99}
\bibitem{footnote} Each member of the hierarchy of equations for powers of
$\eta$ resulting from (\ref{Hyp_eqn}) possesses a measure-valued
traveling wave solution similar to (\ref{M-puls.def}). However, we shall
concentrate on solutions of this type for equation (\ref{EPeqn}).
\end{thebibliography}
\end{document}